\documentclass{aa}
\usepackage[]{natbib,amssymb}
\usepackage{graphicx}
\bibpunct{(}{)}{;}{a}{}{,}
\begin{document}
\title{The signature of substructure on gravitational lensing in
  the $\Lambda$CDM cosmological model}

   \author{M. Brada\v{c} \inst{1,2} \and P. Schneider \inst{1}
    \and M. Lombardi \inst{1,3} \and M. Steinmetz \inst{4,5} \and
    L.V.E. Koopmans \inst{6,7} \and Julio F. Navarro \inst{8}} 
   \offprints{Maru\v{s}a Brada\v{c}}
   \mail{marusa@astro.uni-bonn.de}
   \institute{Institut f\"{u}r Astrophysik und Extraterrestrische
              Forschung, Auf dem H\"ugel 71, D-53121 Bonn, Germany
    \and Max-Planck-Institut f\"{u}r Radioastronomie, Auf dem
   H\"{u}gel 69, D-53121 Bonn, Germany
    \and European Southern Observatory, Karl-Schwarzschild
   Str. 2, D-85748 Garching bei M\"{u}nchen, Germany
     \and Astrophysikalisches Institut Potsdam, An der Sternwarte 16,
    D-14482 Potsdam, Germany
    \and Steward Observatory, 933 North Cherry Avenue, Tucson, AZ
   85721, USA
    \and Kapteyn Astronomical Institute, P.O.Box 800, 9700AV Groningen, The 
Netherlands
    \and Space Telescope Science Institute, 3700 San Martin Drive, 
        Baltimore, MD 21218, USA  
        \and Department of Physics and Astronomy, University of
    Victoria, BC V8P 1A1, Canada
}
\date{Submitted to A\&A}
%
%

\abstract{We present a study
of the lens properties of quadruply imaged systems,
lensed by numerically simulated galaxies.  We investigate 
a simulated elliptical and disc galaxy drawn from high
resolution simulations of galaxy formation in a concordance 
$\Lambda$CDM universe. The simulations include the effects of gas
dynamics, star formation and feedback processes. 
Flux-ratio anomalies observed in strong gravitational lensing 
potentially 
provide an indicator for the presence of
mass substructure in lens galaxies as predicted from CDM simulations. 
We particularly concentrate on the
prediction that, for an ideal cusp caustic, the sum of the signed
magnifications of the three highly magnified images should vanish when the 
source approaches the cusp. Strong violation of this cusp relation
indicates the presence of substructure, regardless of the global,
smooth mass model of the lens galaxy. We draw the following
conclusions: (1) the level of substructure present in simulations produces
violations of the cusp relation comparable to those observed, (2) 
higher-order catastrophes (e.g. swallowtails) 
can also cause changes of the order of 0.6 in the cusp relation 
as predicted by a 
smooth model, (3) the flux anomaly
distribution depends on the image parity and flux and both 
the brightest minimum and saddle-point images are more affected by
substructure than the fainter images.  In addition, the brightest 
saddle point is demagnified w.r.t. the brightest
minimum. Our results are fully 
numerical and properly include all mass scales, without making 
semi-analytic assumptions. They are ultimately limited by the mass 
resolution of single particles in the simulation determined by 
current computational limits, however show that our results are not
  affected by shot-noise due to the finite number of particles.
}

\maketitle
%
%

\def\g{\gamma_1}
\def\gg{\gamma_2}
\def\eck#1{\left\lbrack #1 \right\rbrack}
\def\eckk#1{\bigl[ #1 \bigr]}
\def\rund#1{\left( #1 \right)}
\def\abs#1{\left\vert #1 \right\vert}
\def\wave#1{\left\lbrace #1 \right\rbrace}
\def\ave#1{\left\langle #1 \right\rangle}

\def\vc#1{%
  \if\alpha#1\mathchoice%
    {\mbox{\boldmath$\displaystyle#1$}}%
    {\mbox{\boldmath$\textstyle#1$}}%
    {\mbox{\boldmath$\scriptstyle#1$}}%
    {\mbox{\boldmath$\scriptscriptstyle#1$}}%
  \else
    \textbf{\textit{#1}}%
  \fi}

\newcommand{\newa}[0]{NewA}

%
%
\section{Introduction \label{sc:intro}}
Whereas the current Cold Dark Matter (CDM) paradigm for structure formation is
widely accepted, two major problems for CDM remain. 
While simulations predict
cuspy dark matter halos \citep[e.g.][]{moore94}, 
observed rotation curves of low surface
brightness galaxies indicate that their dark matter halos have more
shallow cores
\citep{kravtsov98,swaters00,vdBosch01,deBlok02}. 
The other is the apparent over-prediction of the small-scale power in CDM
simulations. As was shown by \citet{moore99} and \citet{klypin99},  
the number of satellite halos seen in
N-body simulations appears to far exceed the number of dwarf galaxies
observed around the Milky Way. Particular discrepancies have been found
for satellite masses $\lesssim 10^9 M_{\odot}$.


Gravitational lensing is
at present the only tool to investigate CDM substructure in
galaxies outside the local group. As first noted
by \citet{mao98}, mass-substructure other than stars on scales less than the
image separation can substantially
affect the observed flux ratios in strong gravitational lens systems. 
\citet{chiba02}, \citet{dalal02},  \citet{mao98},
\citet{metcalf01}, \citet{metcalf02}, \citet{metcalf03}, \citet{keeton01},
and \citet{bradac02} have all argued that substructure can 
provide the explanation for
the flux anomalies in various systems. \citet{dalal02} further conclude
 that the amount of substructure needed to explain the flux
ratios of quadruply-imaged systems broadly agrees with the CDM 
predictions. At least for some systems the flux mismatches are
probably not just an artifact of oversimplified macromodels of the
main lens galaxy (see e.g. \citealt{evans02, metcalf02}). 
As discussed by \citet{keeton03} and \citet{chen03}, fluxes can be
further affected by
clumps of matter at a redshift different from that of the lens, 
along the line of sight between the observer and the
source; however, this effect is not dominant. It is also possible that
the small scale structure does not consist of compact CDM clumps, 
also tidal streams or offset disc components can affect the
flux ratios (see \citealt{moeller02,quadri02}).

\citet{keeton01} and \citet{gaudi02} recently focused on the
magnification relations
that should be satisfied by particular four-image geometries
(so called ``fold'' and ``cusp'' configurations). These relations
are model-independent predictions for the magnifications of highly
magnified images \citep{blandford86, blandford90, we92, ma92}.
Strictly speaking, however, they hold
only for ideal ``fold'' or ``cusp'' configurations and 
it is therefore in some cases hard to
disentangle the effects of the source being further away from the cusp
from the effect of substructure, purely by employing these relations.

The influence of substructure can not only be seen on image flux
ratios, but also in the structure of multiple-imaged jets. The lens system
B1152+199 consists of a doubly-imaged jet, one of which appears
bent, whereas the other is not \citep{metcalf02b}. Alternative 
explanation is that an
intrinsic bend in the jet is simply magnified in one image, and
produces only a small effect in the other.

Microlensing can
change the flux ratios not only in the optical (e.g. \citealt{wozniak00}),
but also at radio wavelengths \citep{koopmans00}. Flux ratio anomalies can also be introduced by 
propagation effects
in the interstellar medium (ISM) in the lens galaxy, by galactic scintillation, and scatter broadening \citep{koopmans03b}. 
Fortunately, these effects are frequency dependent and one can
distinguish them using multi-frequency observations. In addition, 
these electromagnetic phenomena are similar for images of
different parities.

For substructure, on the other hand, \citet{schechter02} found that 
magnification perturbations should show a dependence on image parity.
Microlensing simulations showed that the probability
distributions for magnifications of individual images are not 
symmetric around the unperturbed magnification. The distribution 
depends on image parity and becomes highly skewed. The probability
for the brightest saddle point
image  to be demagnified is increased.\footnote{Images form at extrema of the arrival time surface
  (i.e. Fermat's principle). At saddle points, images have negative
and at minima/maxima, they have positive parity. In quadruply-imaged
system one observes two saddle point images, and two minima. The fifth
image is a maximum, too faint to be observed.}  
Observed lens 
systems also seem to show this image parity
dependence \citep{kochanek03}, and this indicates that
the flux ratio anomalies arise mainly from gravitational lensing, rather than
propagation effects.

All these effects on flux ratios have placed some doubt as to 
whether the existence of 
substructure can be rigorously tested with strong lensing and what
is expected signal. Several groups have tested
the effects of substructure in strong lensing systems using a semi-analytic prescription for substructure
\citep{metcalf01, keeton02, dalal02, kochanek03}. Recently
\citet{mao04} showed using high-resolution numerical simulations that
the fraction of surface mass density in substructures is lower 
than required by lensing; however both predictions are still uncertain.

In this paper we use different projections of two different galaxies
obtained in N-body+gasdynamics simulations. Whereas a semi-analytic 
prescription overcomes the problem of shot-noise, and
the problem of modelling becomes
simpler (one has an analytic model for the underlying
macro-distribution), by using the direct output of an N-body simulated
galaxy, one does not make any simplifying assumption about the mass profiles of
the macro model, or the substructure. Down to the resolution scales
of the simulation we therefore believe we have a better comparison with
a realistic galaxy.

Whereas higher-resolution DM-only simulations are available, the
absence of baryons significantly affects lens properties. Those type
of simulations are of limited use for the purpose of testing CDM
substructure effect on strong lens systems. More precise, 
strong gravitational lensing is probing the galaxy potential on
the inner $5-10\mbox{ kpc}$ (the typical size of the
Einstein radius). It was shown by \citet{treu04}
that the range of projected baryonic fraction within the Einstein
radius is $f_{\rm bar} = 0.3-0.6$. It is therefore crucial to
include the gravitational pull of baryons in our simulations.

This paper is structured as follows. In Sect.~\ref{sc:nbody} we
first give the main properties of the N-body simulations that we use. 
We also introduce an improved smoothing scheme compared to
\citet{bradac02} and describe how to extract lensing
properties from N-body simulations. In Sect.~\ref{sc:cusprel} we
focus on the cusp relation for simulated lens
systems. Sect.~\ref{sc:models} describes the modelling of synthetic
images and the phenomenon of suppressed saddle points. We conclude and
give an outlook in Sect.~\ref{sc:conclusion}.

\section{Strong lensing by a simulated galaxy \label{sc:nbody}}

N-body simulations can provide a powerful benchmark for testing the
effects of substructure on strong lensing. One can simulate
conditions in which propagation effects due to the ISM can be ignored and
thus examine only the signature of substructure. The drawback of this
method at present lies in the resolution available for simulations 
that include dark matter, gas and star particles. 
This limits our analysis to mass clumps of
$\gtrsim 10^8 M_{\odot}$. However, since the mass resolution is improving
rapidly, this will soon be less of a problem.

As in \citet{bradac02}, we used the nummerical N-body
simulations for several realisations of galaxies 
including gas-dynamics and star formation
\citep{st01}. 
We investigate two different halos, each of them in three
different projections. The simulations were performed using GRAPESPH,
a code that combines the hardware N-body integrator GRAPE with the
Smooth Particle Hydrodynamics (SPH) technique \citep{st96}.

\begin{table}[b!]
\caption{Properties of the two simulated halos we used. $z_{\rm l}$
denotes the redshift of the halo, $z_{\rm s}$ is the redshift of the
source. $N_{\rm bar}$, $N_{\rm DM}$, and $N_{\rm str}$ are the numbers
of baryonic, dark matter particles and ``stars'', respectively, present
in the cut-out of the simulation we used (note that even within one
family particles have different masses). $M_{\rm tot}$ is the total
mass of the particles we used.}
\label{tab:halos}
\begin{center}
\begin{tabular}{p{0.1\linewidth}c c}
\noalign{\smallskip}
\hline
\noalign{\smallskip}
\hline
\noalign{\smallskip}
Halo & Elliptical & Disk\\
\noalign{\smallskip}
\hline
\noalign{\smallskip}
$z_{\rm l}$ & 0.81 & 0.33 \\
$z_{\rm s}$ & 3.0 & 3.0 \\
$N_{\rm bar}$ & 12\,000 & 20\,000\\
$N_{\rm DM}$ & 17\,000 & 26\,000\\
$N_{\rm str}$ & 70\,000 & 110\,000\\
$M_{\rm tot}$ & $1.5 \times 10^{12} M_{\odot}$ & $0.5 \times 10^{12} M_{\odot}$\\
\noalign{\smallskip}
\hline
\end{tabular}
\end{center}
\end{table} 
In Table~\ref{tab:halos} the properties of the
halos are listed. In both cases  the original simulated field
contains approximately 300\,000  particles. The simulation is 
contained within a
sphere of diameter $32\; {\rm Mpc}$ which is split into a
high-resolution sphere of diameter  $2.5\;{\rm Mpc}$ centred around the
galaxy and an outer low-resolution shell. Gasdynamics and star
formation are restricted to the high-resolution sphere,
while the dark matter particles of the low-resolution sphere sample the
large scale matter distribution in order to appropriately reproduce the
large scale tidal fields (see \citealt{na97} and \citealt{st00} for details on
this simulation technique). From the original simulated field we use a
cube of size 
$\sim 200^3 \: {\rm kpc}^3$ centred on a single
galaxy. This volume lies well within the high-resolution sphere and is
void of any massive intruder particles from the low-resolution shell.

All simulations were performed in a $\Lambda$CDM cosmology
($\Omega_0=0.3$, $\Omega_{\Lambda}=0.7$, $\Omega_{\rm b}=0.019/h^2$,
$\sigma_8=0.9$).  They have a mass resolution of $1.26 \times 10^{7}
M_{\odot}$ and a spatial resolution of $0.5 \: {\rm kpc}$. A
realistic resolution scale for an identified substructure is
typically assumed to be $\sim 40$ particles which corresponds to $5
\times 10^{8} M_{\odot}$. The quoted mass resolution holds for
gas/stars. The high-resolution dark matter particles are about a
factor of 7 ($=\Omega_{0}/\Omega_{\rm b}$) more massive. A detailed
analysis of the photometric and dynamical properties of the simulated
halos was carried out in \citet{meza03} for the elliptical
and \citet{abadi02, abadi02b} for the disc galaxy.

Early N-body 
simulations suffered from the problem of overmerging, i.e. most
satellites were dissolved due to the tidal field of
the host galaxy. Current N-body simulations demonstrate that quite
often this tidal disruption is inefficient resulting in surviving
substructures. While this qualitatively different result is usually
accounted to the increased particle numbers of modern
simulations, it is in fact rather caused by a more prudent choice of
the numerical softening compared with a more rigorous limit on
the numerical time stepping. Convergence studies using modern N-body 
simulations demonstrate that with a sufficiently small
numerical softening length and sufficiently rigorous numerical time 
stepping, mass functions can be accurately produced down to
clumps with only a very few tens of particles \citep[see e.g.][]{springel01}.

\subsection{Delaunay triangulation smoothing technique \label{sc:delaunay}}

From the irregularly sampled particle distribution  in the simulation
box, we reconstruct the density field. We apply a 
smoothing procedure and then project the
resulting particle distribution to obtain the surface mass density
$\kappa$. In \citet{bradac02} it was shown that
a more sophisticated smoothing method should be employed for the data
analysis than simply smoothing with a Gaussian
kernel. A method is needed that adapts the kernel size in order to
increase the signal to noise of the reconstructed field.  
For this purpose, we make use of the Delaunay tesselation
technique from \citet{schaap00}.

We perform a three-dimensional Delaunay
triangulation using the QHULL algorithm \citep{barber96}. The density
estimator from \citet{schaap00} is then evaluated at
each vertex, and we 
interpolate values of the density at each three-dimensional grid
point to obtain the $\kappa$ map.  The resulting
density field is then projected onto a
two-dimensional grid.

Since the N-body simulations contain three independent
classes of particles (gas, stars, and dark matter
particles, each having different masses), we repeat the
procedure described above for each group separately and the 
final $\kappa$ map is obtained by adding the contributions from all three classes.

The Delaunay tesselation method performs very well in comparison to
the standard Gaussian
smoothing technique used in \citet{bradac02}, or an adaptive Gaussian
smoothing technique (see also \citealt{schaap00}). Since it is
non-parametric, it adjusts the scale of the
smoothing kernel such that regions of low noise
(i.e. where the particle density is highest) are effectively smoothed 
less than regions with high noise. Also the shape of
the kernel is self-adaptive. Hence, this method is very useful for
the analysis of galaxies with high dynamic range and significant structure
in the mass distribution (e.g. mass clumps and spiral arms).    
 
\subsection{Estimating the noise properties \label{sc:delaunay_noise}}
  
A drawback of using the Delaunay tesselation method is that the
signal-to-noise evaluation for the final surface mass density map is
non-trivial.  For example, with a Gaussian kernel one can determine 
the noise by simply looking at the number
of particles in a smoothing element (for a more detailed estimate, see
\citealt{lombardi02}).  
When using the tesselation technique, such an approach is not viable.

One approach for 
estimating the error is to use bootstrapping (see
e.g. \citealt{heyl94}). Normally we calculate physical properties
by using all $n$ particles in the simulation. 
To create a bootstrap image one has to randomly select $n$
particles out of this simulation with replacement; i.e. some of the
particles from the original simulation will be included more than
once and some not at all. In other words, we randomly generate $n$
integers from 1 to $n$, representing the bootstrapped set of
particles. If a particle is included $k$ times in a bootstrapped
map, we put a particle at the same position with $k$ times its
original mass. One can
then make an error estimation using the ensemble of such images and 
calculating the desired physical quantity for each of them.

Whereas the tesselation itself is done very quickly, 
the interpolation of density on a grid is a process that
takes a few CPU days on a regular PC. We therefore limit
ourselves to 10 bootstrapped
maps and the elliptical halo only.

For each pixel $i$ we calculate the associated error using
\begin{equation}
\sigma^2\rund{\kappa_i} = \frac{1}{M-1}\sum_{m=1}^{M} \rund{\kappa^{m}_i - \ave{\kappa_i}}^2
\label{eq:var}
\end{equation}
where $\ave{X_i}$ is the average value of $X$ at pixel $i$ averaged
over all $M$ bootstrapped maps; in our case $M = 10$. 
This procedure gives us an estimate for the error on
$\kappa$. We find that within the critical curves, 
i.e. $\kappa \gtrsim 0.25$
the average noise $\sigma(\kappa_i) / \ave{\kappa}$ is of the order 
of $\lesssim 5\%$.

\subsection{Strong lensing properties \label{sc:lensprop}}
\begin{figure}[ht!]
\begin{center}
\includegraphics[width=7.5cm]{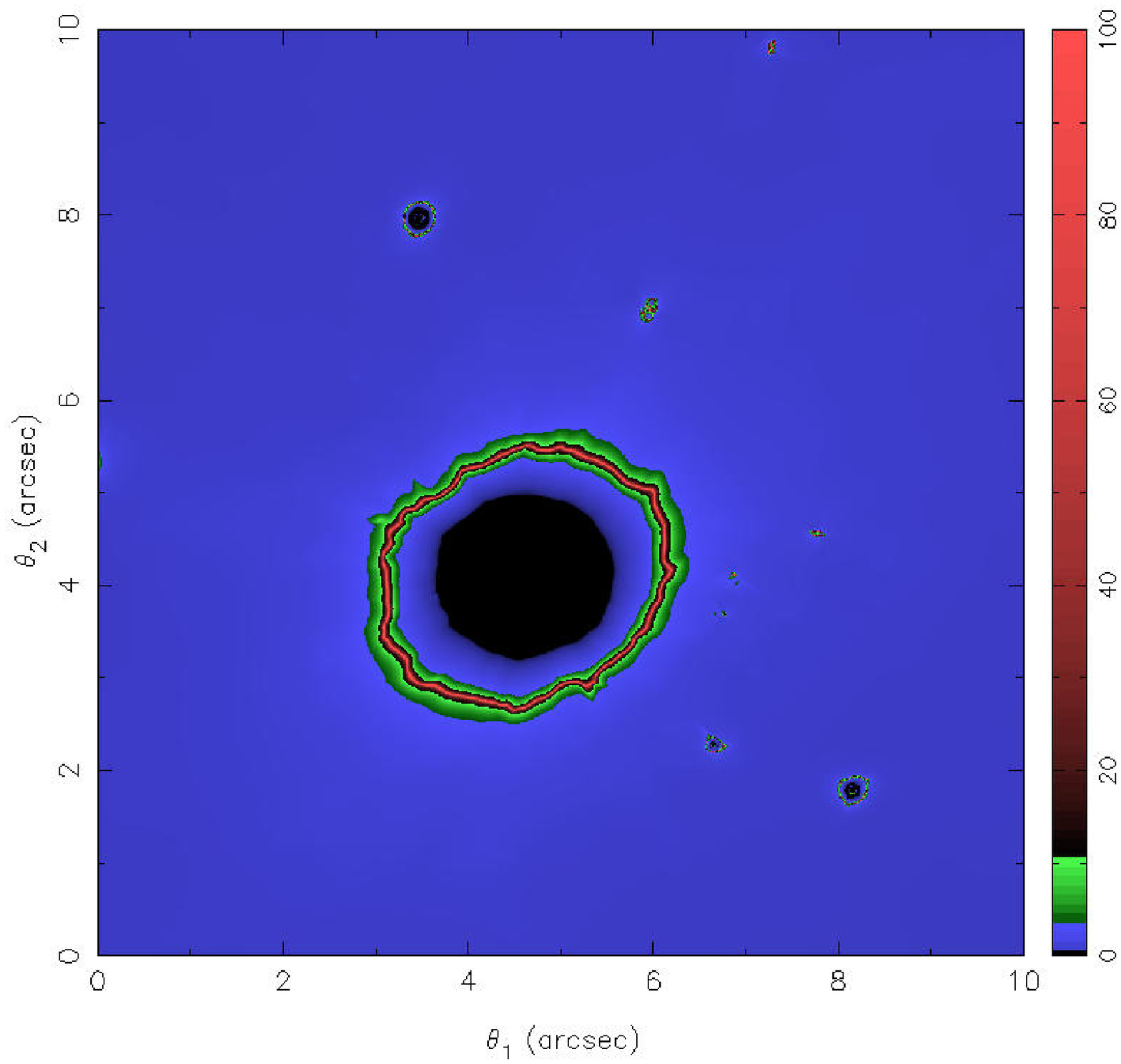}
\centerline{{\bf (a)}}
\vspace{0.2cm}
\includegraphics[width=7.5cm]{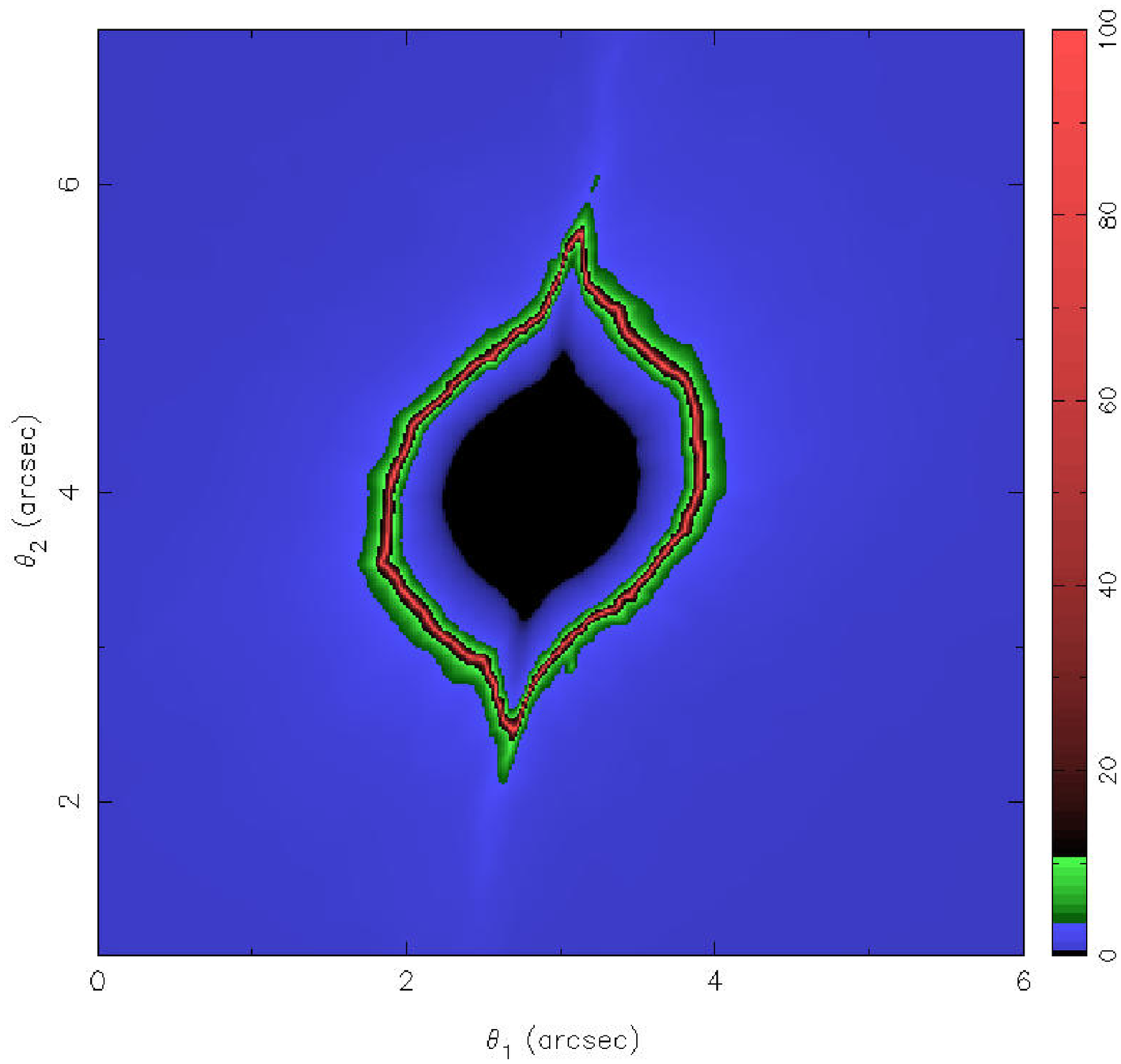}
\centerline{{\bf (b)}}
\end{center}
\caption{The magnification map of the simulated elliptical {\bf (a)} and
edge-on disk {\bf (b)} galaxy. 
External shear is added in the evaluation of the
magnification map to account for neighbouring galaxies (see
text). Lighter regions represent high magnifications. The units on the
axes are arcseconds, one arcsecond in the {\it lens} plane corresponds to
approximately $7 \:{\rm kpc}$ in {\bf (a)} and   $5 \:{\rm kpc}$ in
{\bf (b)}.\protect\footnotemark}
\label{fig:magnification}
\end{figure}\footnotetext{For distance calculations 
throughout the paper we assume an Einstein-de-Sitter
Universe and the Hubble constant $H_{0}=65\:{\rm km\, s^{-1}\,
Mpc^{-1}}$.}
Having obtained the $\kappa$-map, we then calculate the lens
properties on the grid ($2048 \:\times \:2048$ pixels). 
The Poisson equation for the lens potential $\psi$
\begin{equation}
\nabla^2 \psi(\vc \theta)=2\, \kappa(\vc \theta)
\label{eq:t.16}
\end{equation}
is solved on the grid in Fourier space with a DFT (Discrete Fourier
Transformation) method. We used the publicly
available C library FFTW (``Fastest Fourier Transform in the West'')
written by \citet{frigo98}. To reduce the boundary effects,
padding was introduced. In particular, the  
DFT was performed on a $4096 \:\times \:4096$ grid, where one fourth
of the grid contains the original $\kappa$ map and the rest is set to
zero.

One can now proceed in two ways. Either, one can calculate
the two components of the shear $\gamma_{1,2}$ and the deflection angle $\vc
\alpha$ by multiplying the potential in Fourier
space by the appropriate kernel. However, since we calculate the Fourier transform of
the potential on a finite grid, we filter out high
spatial frequency modes. By multiplying the transform with different 
kernels for
$\gamma_{1,2}$ and  $\vc \alpha$, these final maps
do not correspond exactly to the same $\kappa$ map. The effect is
small, but it shows up near the critical curves.
Therefore, it is better to only calculate the lens potential $\psi$ 
using DFT and obtain the shear and deflection angle by finite
differencing. The latter method is also less CPU-time and
memory intensive.

The simulated galaxies are a field galaxy. However, most of the lenses in 
quadruple
image systems are members of groups. To make our simulated
galaxies more closely resemble a realistic system, we add 
an external shear to the Jacobi matrix 
(evaluated at each grid point). The shear components are the same for
each projection and all halos. We use
\[
\gamma_1^{\rm ext}=-0.04 \; , \quad \gamma_2^{\rm ext}=-0.16 \; ;
\]
corresponding to the shear of the best-fit singular-isothermal
ellipse model with external shear for the lens B1422$+$231 in 
\citet{bradac02}. 
Figure~\ref{fig:magnification} show the magnification maps of
the elliptical and edge-on disk galaxy.
The corresponding caustic curves (for a source at $z=3$) were
obtained by projecting the points with vanishing  determinant of the
Jacobian matrix from the image
plane to the source plane. The
  critical curves are
plotted in white  in Fig.~\ref{fig:magnification}a for the elliptical and
Fig.~\ref{fig:magnification}b for the edge-on
disk galaxy. The corresponding caustic curves are plotted in
Fig.~\ref{fig:cusprel}a and \ref{fig:cusprel_s}c respectively.

\subsection{The importance of baryons \label{sc:dmvssph}}
\begin{figure*}[ht!]
\begin{minipage}{17cm}
\begin{minipage}{8.5cm}
\begin{center}
\includegraphics[width=0.9\textwidth]{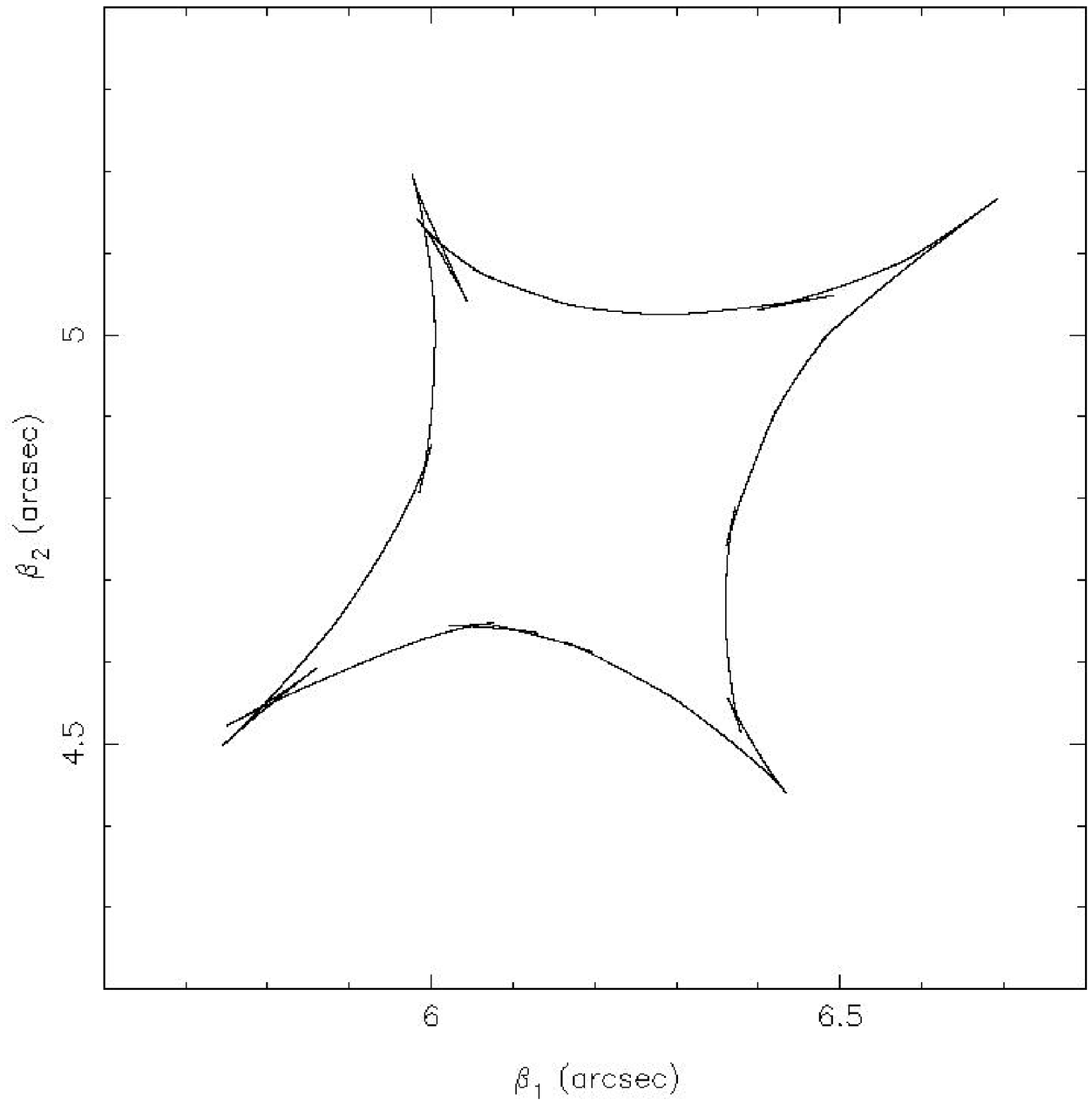}
\centerline{{\bf (a)}}
\end{center}
\end{minipage}
\begin{minipage}{8.5cm}
\begin{center}
\includegraphics[width=0.9\textwidth]{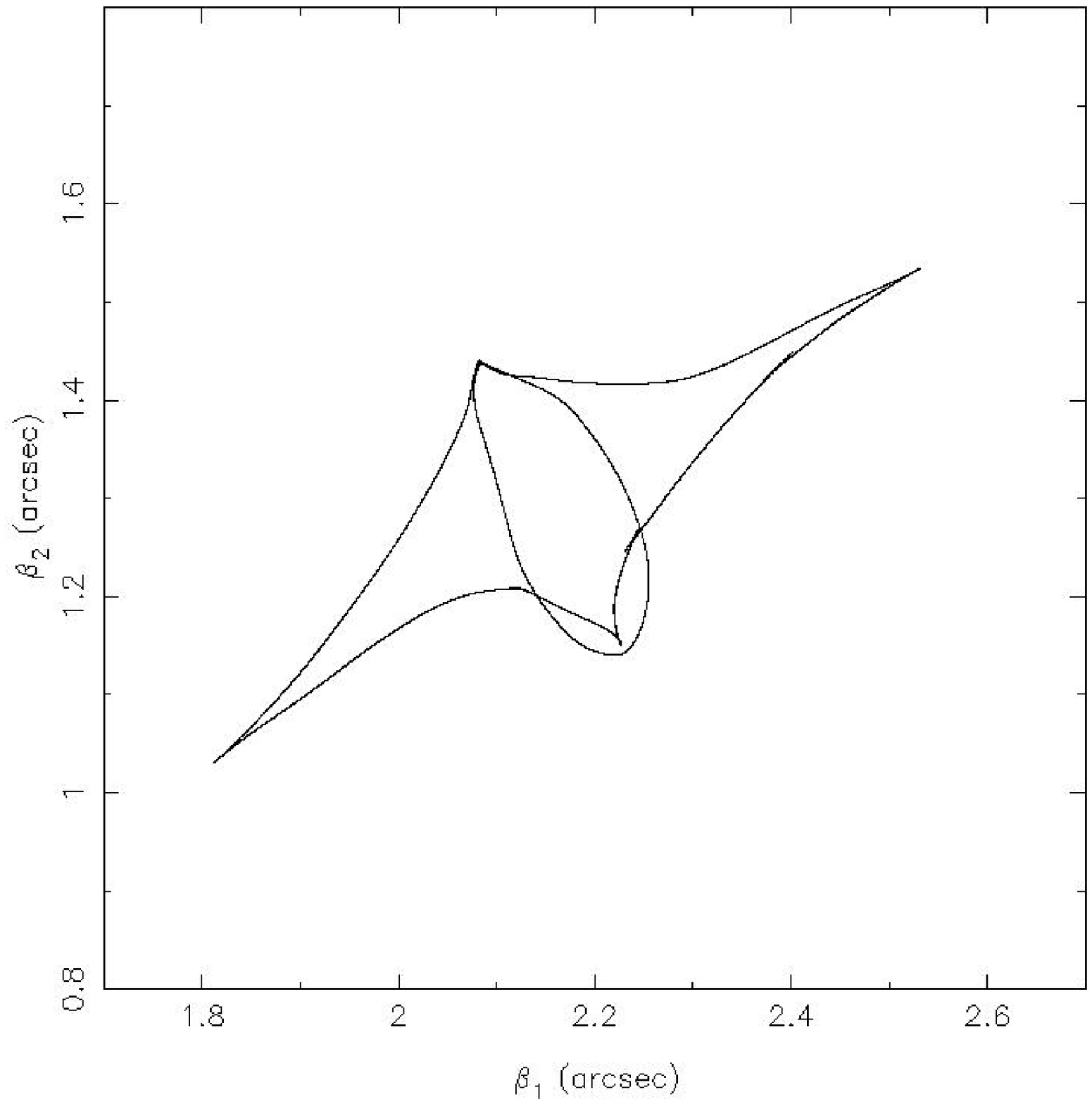}
\centerline{{\bf (b)}}
\end{center}
\end{minipage}
\end{minipage}
\caption{The caustic curves
  of two simulated galaxies. Panel {\bf (a)} represents the simulation
  which includes baryonic and dark matter particles, 
whereas for panel {\bf (b)} we use
a simulation with dark matter particles only. The radial caustic 
for the dark matter only simulation it is almost
  entirely enclosed within the asteroid caustic, prohibiting formation
  of cusp images in quadruply-imaged systems.}
\label{fig:cuustic_dm}
\end{figure*}

As mentioned above, the influence of baryons is
  very important in lens galaxies. Since we are interested in the 
effects of substructure, it would be desirable to use N-body
  simulations that have the highest possible dynamical
range. At present this is achieved in  
high-resolution
N-body simulations that only include dark matter. However, 
if baryons are not included, the central potential is
more shallow than what we typically observe in lens galaxies. 
All quadruply-imaged systems for which the inner
slope of the mass distribution has been measured, are well described
by a total mass density 
profile $\sim r^{-\gamma}$ with $\gamma \simeq 2$ 
\citep{kochanek95,cohn01, treu02, treu02b, koopmans03,treu04},
consistent with the combined mass distribution of dark
matter and baryons seen in the simulations. 
Hence, dark matter only simulations do not
accurately represent the overall properties of lens galaxies, 
and instead we need to use hydro-dynamical simulations.

To test the importance of baryons 
we have simulated an elliptical halo using only dark
matter particles and performed the same lensing analysis as described above. We
project the density approximately along the long axis of the halo,
thus maximising the central density.  In
Fig.~\ref{fig:cuustic_dm} we plot the corresponding caustic curves for
the halo simulations; in panel (a) the simulation including baryons
(DM+B), for panel (b) we use the DM-only simulation (DM). 
The radial critical curve in
the DM+B halo is located close to the galaxy centre and is therefore not well
resolved (but also irrelevant for our purposes) - the corresponding caustic is therefore not plotted.

The two caustic curves are very different, indicating very different
overall strong lensing properties of these two simulations. Whereas
the DM+B simulation has a steep inner mass profile (very close to
singular isothermal), the DM simulation has a
caustic configuration typical for a lens with a shallow density profile
\citep[see e.g.][]{wa93}. The radial critical curve has become smaller
compared 
to the DM+B halo and the corresponding caustic curve is almost
entirely enclosed by
the asteroid caustic. The prominent naked cusp region is a three-image
region. This configuration is
extremely rare among the observed lensed systems. For a source located
in such a region one would observe three highly magnified images. 
There is only one
possible example of a triply imaged quasar out of $\sim 50$
doublets and triplets \citep{evans02b}, namely \mbox{APM
  08275+5255} \citep{ibata99}. Further, the vast majority of systems 
similar to B1422+231 can not form in such a potential.
We therefore conclude that
the effects of baryons have to be included, and we
will use only DM+B simulations from now on, discussing their limits
where necessary.

\section{The cusp relation \label{sc:cusprel}}

We generate different four-image systems using
each simulated galaxy.  For each projection, regions 
in the source plane where five images form are determined. 
The image plane is projected back to the
source plane using the magnification and deflection angle maps. We used
the grid search method from \citet{blandford87} to find the
pixels enclosed by the asteroid caustic and approximate image
positions. Then the MNEWT routine from \citet{numrec_c} is applied
and the lens equation is solved for the image positions. For this step, we
interpolate the deflection angle 
between the grid points. We use
bilinear and bicubic spline interpolation, and both methods give comparable
results. Once we have the image positions, their magnifications are
calculated and the four brightest images are chosen. These represent the
``observable'' images; the fifth image is usually too faint
(except in the regions where more than five images are formed) and
therefore likely to escape observation.

We can now investigate
the basic properties of the synthetic four-image systems.
There are three basic configurations:
the fold, cusp, and cross. They correspond to a source located inside the
asteroid caustic, close to a fold, a cusp or near the centre,
respectively (see e.g. \citealt{keeton01}). All configurations 
have been observed, and even though
one would naively think that fold and cusp images are rare among
observed lenses, they are in fact frequently observed due to the large
magnification bias. In this section we will mainly concentrate on 
cusp image configurations.

The behaviour of gravitational lens mapping near a cusp was 
first studied by
\citet{blandford86}, \citet{blandford90}, \citet{we92}, and
\citet{ma92}, who investigated the magnification properties
of the cusp images and concluded that the sum of the signed
magnification 
factors of three merging images approaches zero as the source moves 
towards the
cusp. In other words,
\begin{equation}
R_{\rm cusp} = \frac{\abs{\mu_{\rm A} + \mu_{\rm B} + \mu_{\rm C}}}
{\abs{\mu_{\rm A}} + \abs{\mu_{\rm B}} + \abs{\mu_{\rm C}}}
\rightarrow 0, \; {\rm for} \; \mu_{\rm tot} \rightarrow \infty\;, 
\label{eq:rcusp}
\end{equation}
where $\mu_{\rm tot}$ is the unsigned sum of magnifications of all
four images, and A, B and C is the triplet of images forming the
smallest opening angle \citep{keeton01}. The opening angle is measured
from the galaxy centre and is spanned by the two images of equal
parity. The third image lies inside this opening angle.

\subsection{The cusp relation of an N-body simulated elliptical galaxy 
\label{sc:cusprel_nbodye}}
\begin{figure*}[ht!]
\begin{minipage}{17cm}
\begin{minipage}{8.5cm}
\begin{center}
\includegraphics[width=0.9\textwidth]{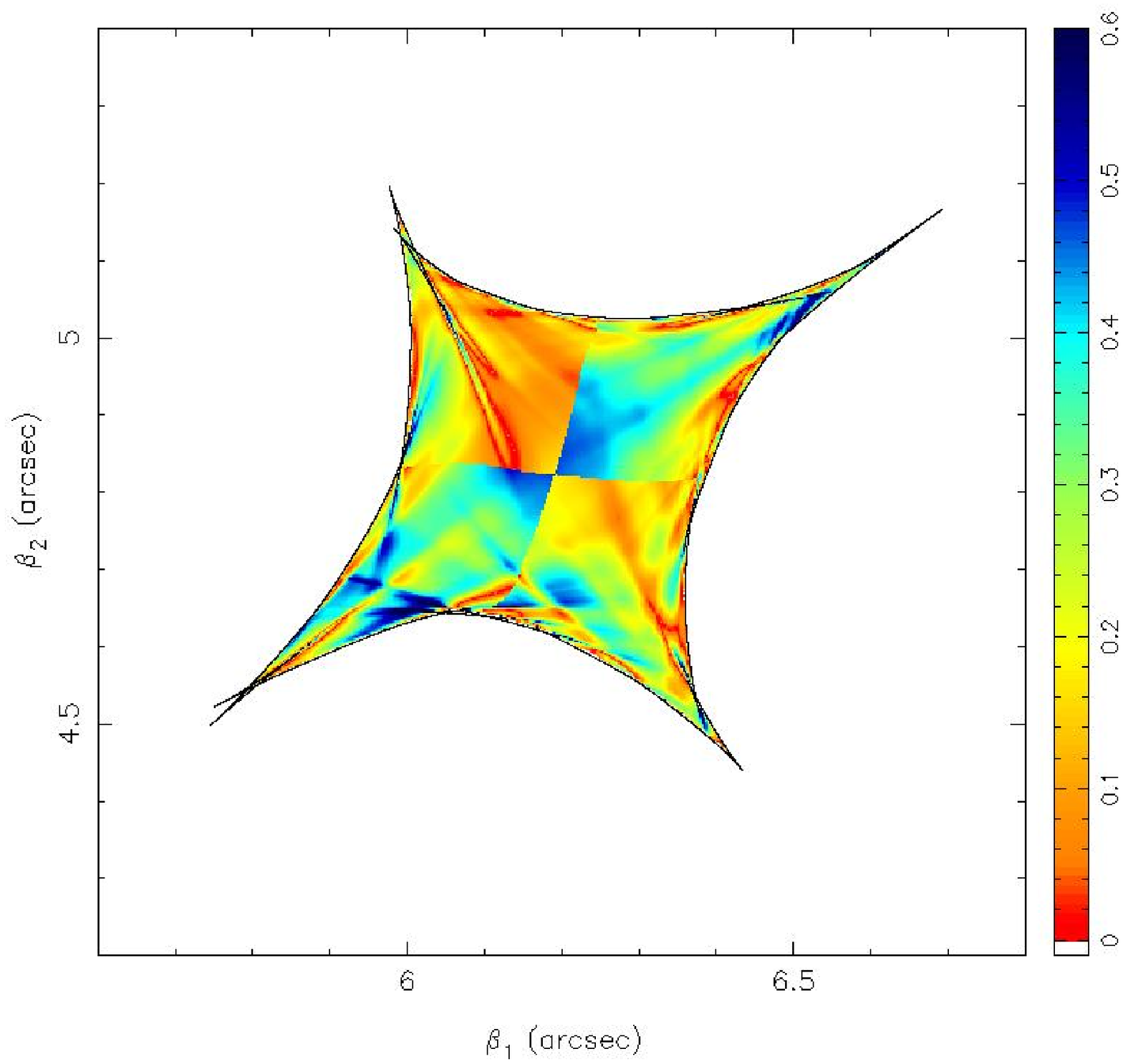}
\centerline{{\bf (a)}}
\includegraphics[width=0.9\textwidth]{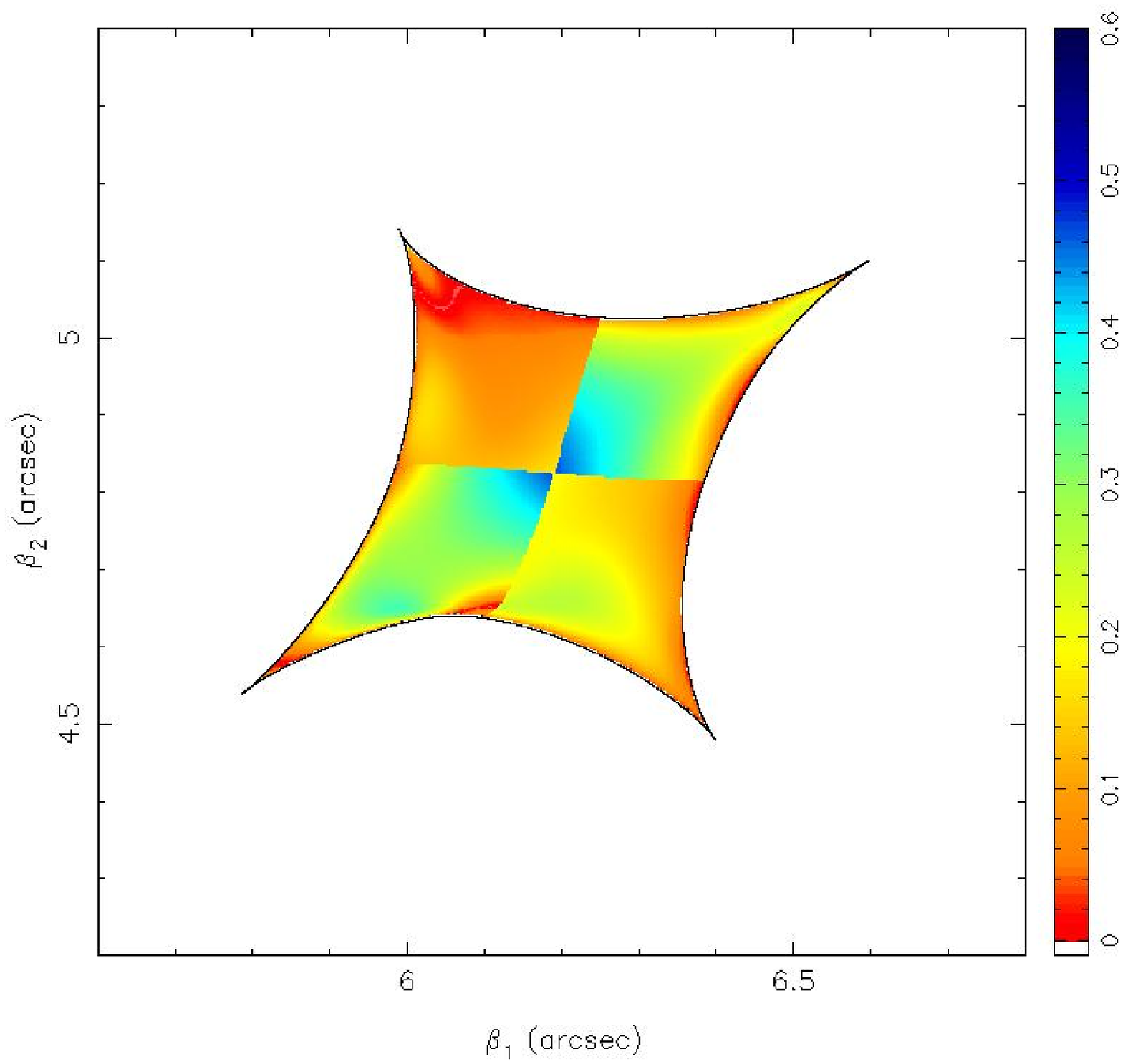}
\centerline{{\bf (c)}}
\end{center}
\end{minipage}
\begin{minipage}{8.5cm}
\begin{center}
\includegraphics[width=0.9\textwidth]{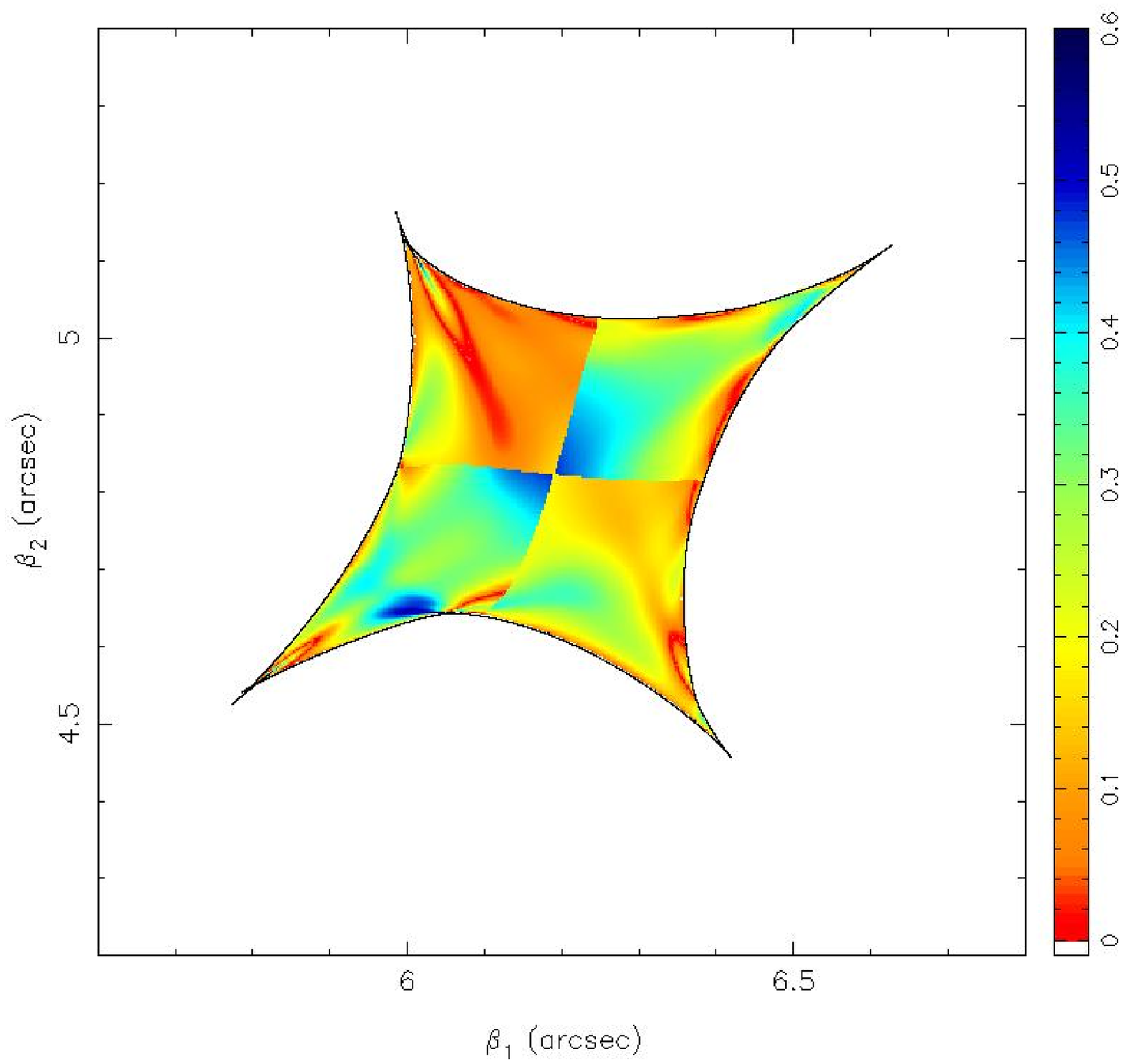}
\centerline{{\bf (b)}}
\includegraphics[width=0.9\textwidth]{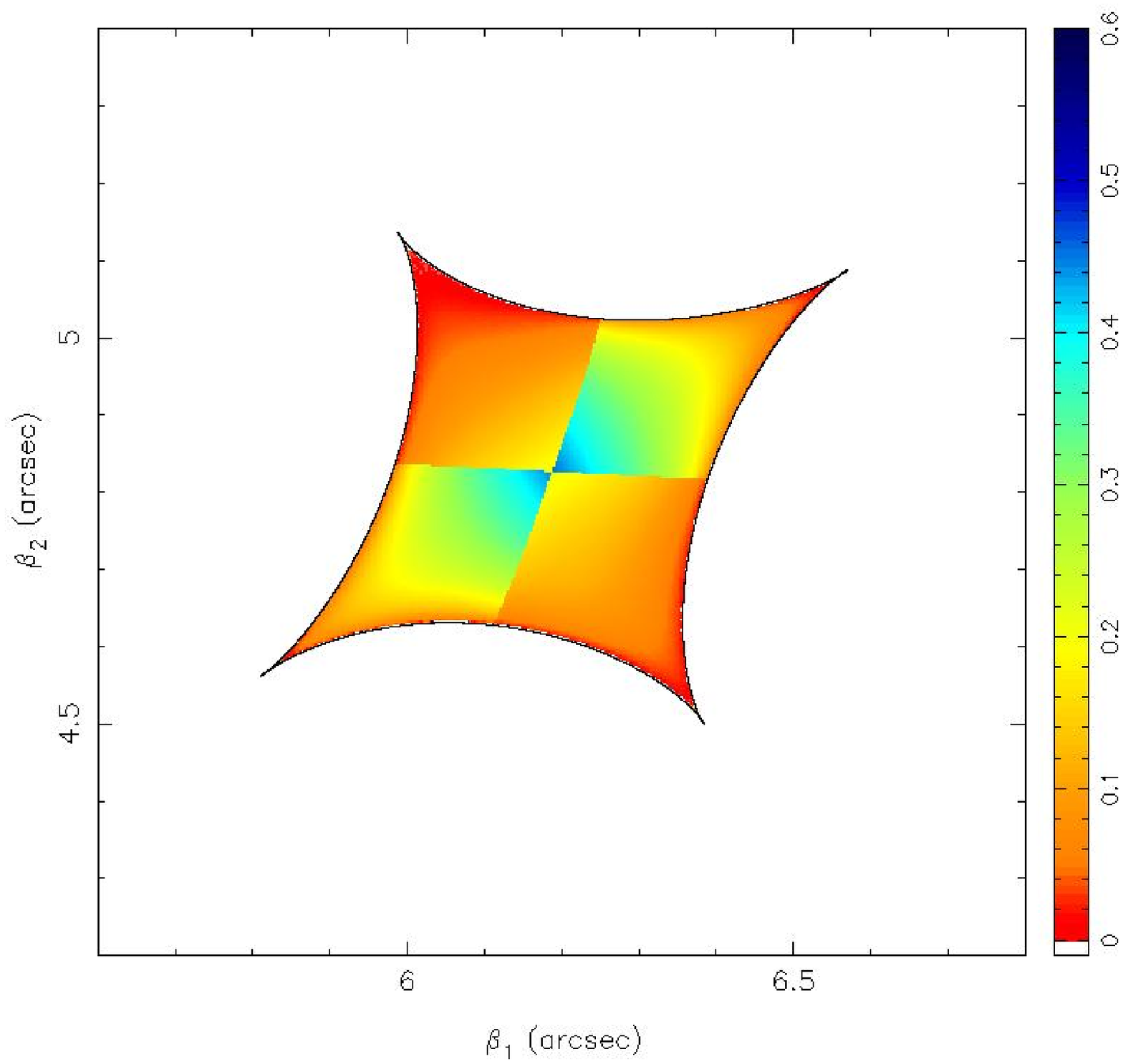}
\centerline{{\bf (d)}}
\end{center}
\end{minipage}
\end{minipage}
\caption{The cusp relation $R_{\rm cusp}$ 
for the N-body simulated elliptical galaxy
at a redshift of $z_{\rm l} = 0.81$. The source was put at a
redshift of $z_{\rm s} = 3$ and approx. 30\,000 systems were generated
that lie inside the asteroid caustic. $R_{\rm cusp}$ is
plotted in gray-scale, for sources close to the cusp the smooth models would
predict $R_{\rm cusp} \sim 0$ (i.e. white - red). The deviations are
due to the substructure. Due to magnification bias most of the observed
lenses correspond to the fold and cusp configurations. Discontinuities
in the maps arise when the source moves in the direction of the 
minor or major axes, since we chose
different subsets of three cusp images. 
On top we plot the caustic curve. Panel {\bf (a)} shows  $R_{\rm cusp}$
for the original mass distribution, 
whereas panels {\bf (b)}--{\bf (d)} show the
cusp relation for the models where we additionally smoothed the
substructure (see text) with a Gaussian kernel characterised by 
standard deviation $\sigma_{\rm G} \sim 1\:{\rm kpc}$ {\bf (b)},
$\sigma_{\rm G} \sim 2\:{\rm kpc}$ {\bf (c)} and  $\sigma_{\rm G} \sim
5\:{\rm kpc}$ {\bf (d)}.} 
\label{fig:cusprel}
\end{figure*}
The cusp relation (\ref{eq:rcusp}) is an asymptotic relation and holds
 when the source approaches the cusp from inside the asteroid. 
One can derive the
 properties of lens mapping close to critical curves using a Taylor
 expansion of the Fermat potential around a critical
 point (see e.g. \citealt{we92}). Such calculations are very cumbersome and 
therefore it is difficult
 (if not impossible) to explore the influence of arbitrary substructure
 analytically. In practice, we can calculate $R_{\rm cusp}$ for the
 N-body simulated systems. Smoothing the original $\kappa$-map on
 different scales then gives an indication of the influence of 
substructure on $R_{\rm cusp}$.

The three cusp images [designated as A, B and C in (\ref{eq:rcusp})]
are chosen according to the image
geometry. Since we know the lens position, this procedure is
straightforward and foolproof. We have identified the triplet of
images belonging to the smallest opening angle (described above).  
Since we know the image parities and
magnifications, one is tempted to identify the three brightest 
images as the cusp images and assign different parity to the brightest
one than to the other two (e.g. as in \citealt{moeller02}). However,
due to the presence of shear and substructure this could lead to 
misidentifications.

Figure~\ref{fig:cusprel}a shows the caustic curve in the source plane
for the simulated elliptical at a redshift of $z_{\rm l} =
0.81$. The source is at a redshift of $z_{\rm s} = 3$.
Approximately 30\,000 lens systems are generated with source 
positions inside the asteroid
caustic. $R_{\rm cusp}$ is plotted in gray-scale.  The apparent 
discontinuities
originate from different image identification. In the very
centre of the caustic the meaning of ``cusp image'' is ill
defined. As the source moves in the direction of the 
minor or major axes we chose
different subsets of three cusp images and therefore the discontinuity
arises.

The remaining panels of Fig.~\ref{fig:cusprel} show the effect of
smoothing the small-scale in the surface mass density $\kappa$ map 
with Gaussian kernels
characterised by standard deviation $\sigma_{\rm G}$. The values for
$\sigma_{\rm G}$ were chosen to be $\sigma_{\rm G} \sim 1,2,5 \; {\rm
kpc}$ for panels (b), (c) and (d) respectively. Note that we do not 
smooth the $\kappa$-map directly.  First we obtain
the smooth model  $\kappa_{\rm smooth}$  for the $\kappa$-map 
by fitting elliptical 
contours to the original map using {\tt IRAF.STSDAS} package
{\tt ellipse}. We subtract $\kappa_{\rm smooth}$ from $\kappa$ and 
smooth the difference using different Gaussian kernels. 
We then add the resulting map back to the $\kappa_{\rm smooth}$. In
this way the overall radial profile of the mass distribution is not
affected.

The effect of smoothing on the cusp relation is clearly visible. In
Fig.~\ref{fig:cusprel}d one sees that the
substructure is completely washed out when smoothing on scales 
of $\sigma_{\rm G} \sim
5\:{\rm kpc}$ is applied. As
we go to smaller smoothing scales,
the effects of substructure become
clearly visible.
In the extensions of swallowtails there is a region where
the cusp relation is strongly
violated (with $R_{\rm cusp}\sim 0.6$, where the smooth model predicts 
 $R_{\rm cusp}\lesssim 0.1$). However, further out, a swallowtail can cause the cusp relation
to change the trend and go to zero (due to high-magnification systems
being formed in such region).

\begin{figure}[ht!]
\begin{center}
\includegraphics[angle = 0,width=7cm]{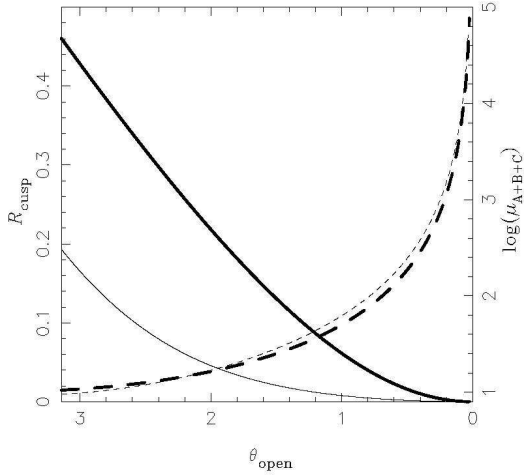}
\end{center}
\caption{$R_{\rm cusp}$ for a simple elliptical isopotential lens
model with $\epsilon = 0.15$. 
$R_{\rm cusp}$ is plotted as a thick (thin) solid line 
for sources along the major (minor) axis. 
$\theta_{\rm open}$ represents the angle measured from the position of
the galaxy, spanned by two ``outer'' cusp images (A and C). The
opening angle $\pi$ means that the source is located at the centre
(images A, C and the galaxy lie on the same line). When the source
approaches the cusp, $\theta_{\rm open} \rightarrow 0$. The total
magnification for the three cusp images $\mu_{\rm A+B+C}$ is plotted 
as a thick (thin) dashed line for sources along the major (minor) axis.} 
\label{fig:cusprel_analytic}
\end{figure}

Finally, the cusp 
relation behaves differently for the source on the major or
the minor axis (see especially Fig.~\ref{fig:cusprel}d). 
This is a generic feature
for smooth elliptical models and can easily be calculated for
e.g. an elliptical isopotential model (see e.g. \citealt{sc92}). We use
this model since it is
analytically tractable for source
positions along the major and minor axis. 
In Fig.~\ref{fig:cusprel_analytic} we plot the cusp relation for the
source moving along the major (minor) axis as a thick (thin) solid line
for the elliptical isopotential model with $\epsilon = 0.15$. As the
source approaches the cusp, $R_{\rm cusp} \rightarrow 0$ for both
source positions, however the slope is different. We also plot the
total magnification factor of the three cusp images, i.e. 
$\mu_{\rm A+B+C} = 
\abs{\mu_{\rm A}} + \abs{\mu_{\rm B}} + \abs{\mu_{\rm C}}$  
as a thick(thin) dashed line.

\begin{figure}[tbp]
\begin{minipage}{8cm}
\begin{center}
\includegraphics[width = 0.8\textwidth]{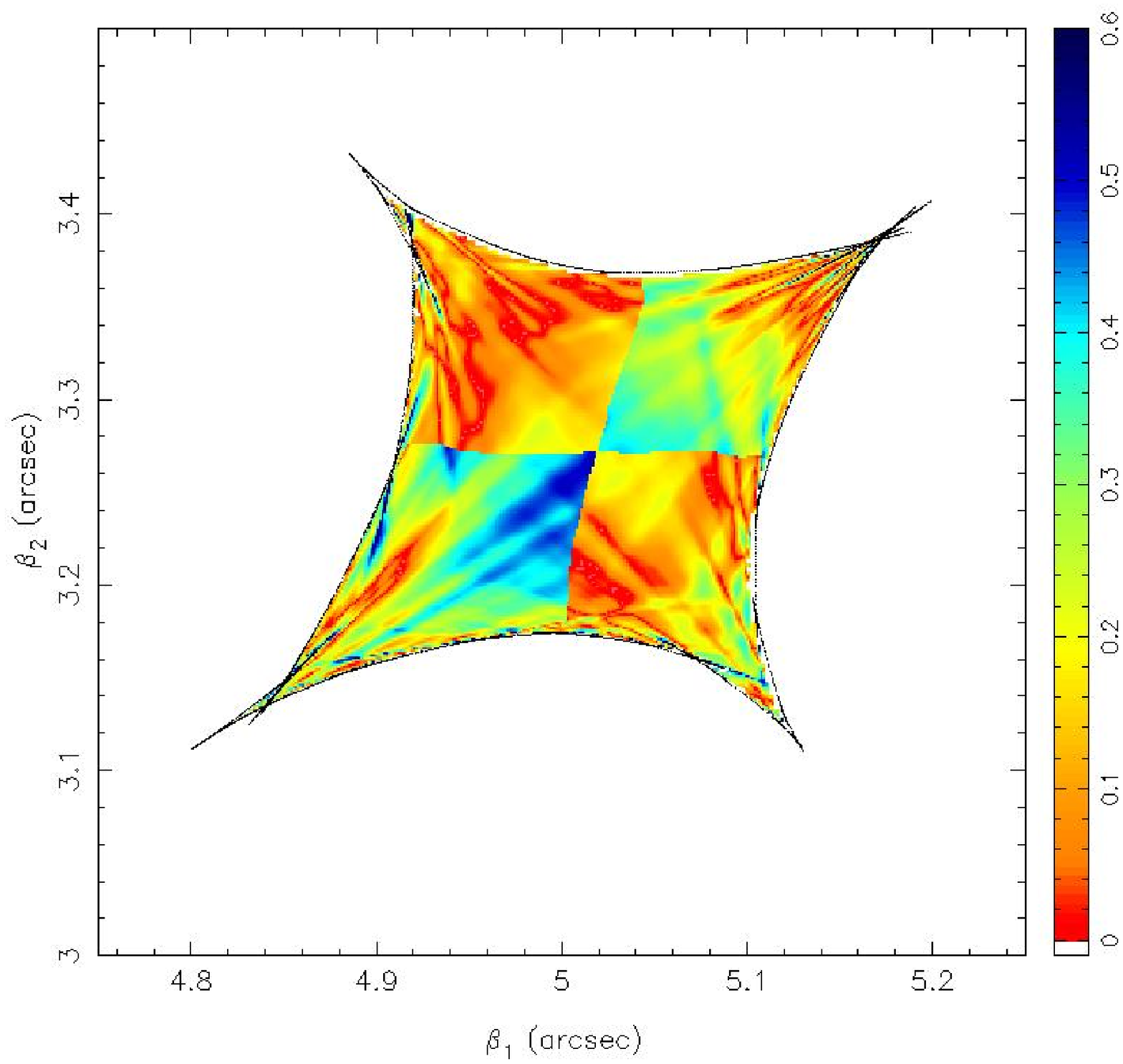}
\centerline{{\bf (a)}}
\includegraphics[width=0.8\textwidth]{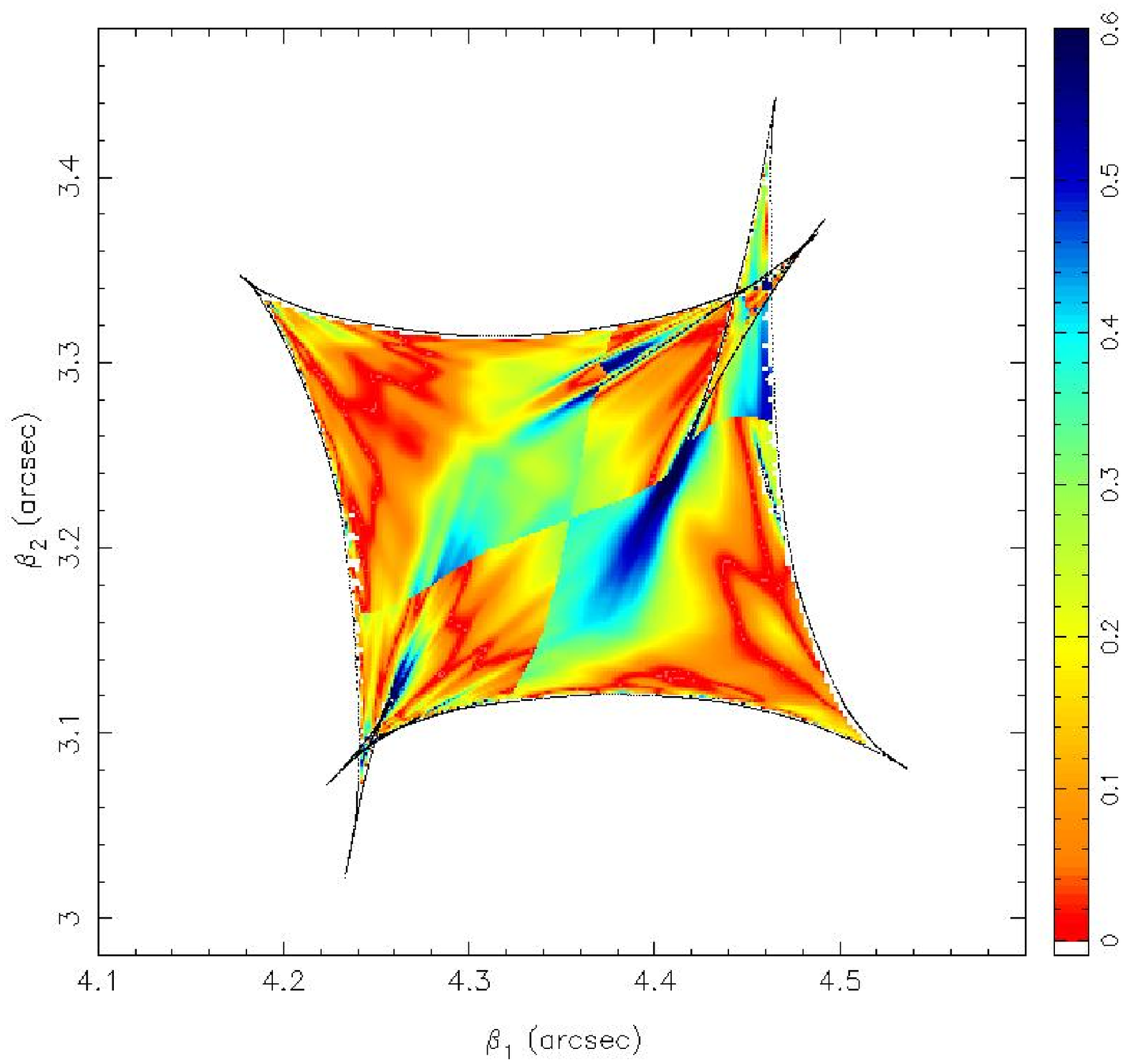}
\centerline{{\bf (b)}}
\includegraphics[width=0.8\textwidth]{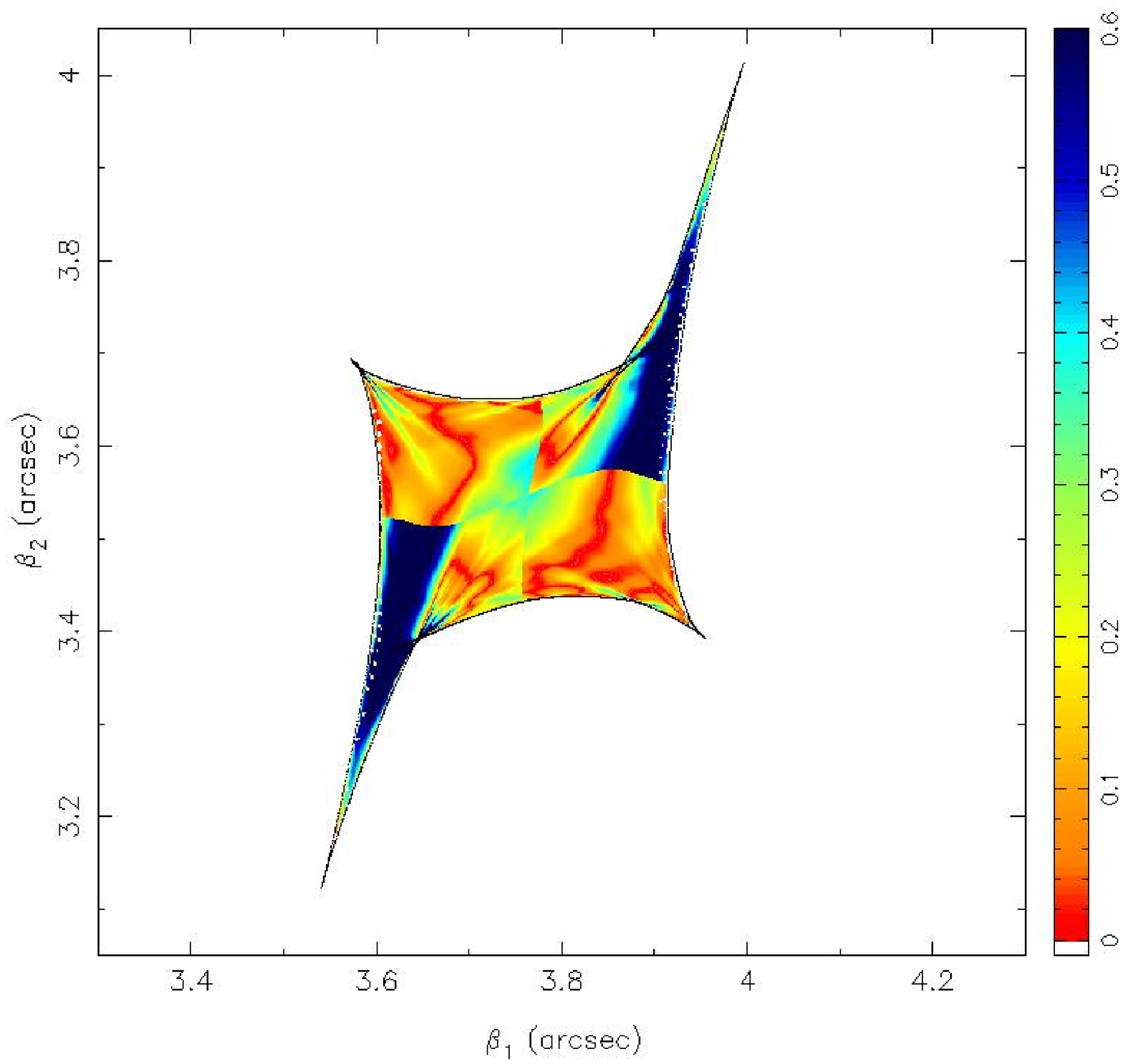}
\centerline{{\bf (c)}}
\end{center}
\end{minipage}
\caption{The cusp relation for the N-body simulated disk galaxy
at a redshift of $z_{\rm l} = 0.33$. The source was put at a
redshift of $z_{\rm s} = 3$ and approx. 10\,000 systems were generated
which lie inside the asteroid caustic. $R_{\rm cusp}$ is
plotted in grey-scale. On top we plot the caustic curve. Panels {\bf (a)},
{\bf (b)}, and {\bf (c)}
show three different orthogonal projections of the halo, {\bf (a)}
corresponds to the  face-on, {\bf (b)} and {\bf (c)} to the edge-on
projection. Panel{\bf (c)}
corresponds to the magnification map in Fig.~\ref{fig:magnification}b.} 
\label{fig:cusprel_s}
\end{figure}

\subsection{The cusp relation in an N-body simulated disk galaxy
\label{sc:cusprel_nbodys}}
The procedure described above was applied to three different projections
of the elliptical halo and very similar conclusions can be
drawn. However, the question
is, how much these conclusions depend on the
specific morphological type of the galaxy. To
investigate this we follow a similar procedure for a
simulated disk galaxy.

In this case, 
however, we did not look at the effects of additional smoothing. 
It is difficult to
subtract a smooth model, since the galaxy consists of a bulge, warped
disk and extended halo, which can not simply be fitted by ellipses. If we
were to smooth the edge-on projection with a Gaussian kernel, we would
also wash out the (in our case warped) disc. 
We only include the analysis of the cusp relation for
this halo in order to show the effect of the disc on $R_{\rm cusp}$.

Figure~\ref{fig:cusprel_s} shows the cusp relation of an N-body
simulated disk galaxy in three projections. The disc 
clearly plays a role for the edge-on projection (see
Fig.~\ref{fig:cusprel_s}b,c), whereas the face-on projection is
similar to the elliptical galaxy. Especially in
Fig.~\ref{fig:cusprel_s}c, where the disc extent in projection
is smaller than the typical image separation, the cusp relation in the
upper-right and lower-left cusp is strongly violated. This direction
corresponds to the orientation of the disc.

\subsection{Observed cusp relation \label{sc:cusprel_obs}}
\begin{table}[b!]
\caption{The values for $R_{\rm cusp}$ taken from \citet{keeton02} 
for four lens systems showing a typical cusp geometry.}
\label{tab:obscusp}
\begin{center}
\begin{tabular}{p{0.5\linewidth}c}
\noalign{\smallskip}
\hline
\noalign{\smallskip}
\hline
\noalign{\smallskip}
Lens & $R_{\rm cusp}$\\
\noalign{\smallskip}
\hline
\noalign{\smallskip}
B2045+265 & $0.52 \pm 0.04$\\
B0712+472 & $0.26\pm 0.06$\\
RX J0911+0551 & $0.22\pm 0.06$\\
B1422+231 & $0.18 \pm 0.06$\\ 
\noalign{\smallskip}
\hline
\end{tabular}
\end{center}
\end{table}

In this section we compare 
predictions of  $R_{\rm cusp}$ from our simulated galaxies with the
values from observed lens system. Unfortunately the number of
observed systems is not large; seventeen four-image systems have been 
published (see \citealt{keeton02} for a summary), 
and four of them are believed to
show a typical cusp geometry (see Table~\ref{tab:obscusp}).

Note that relation (\ref{eq:rcusp}) is model independent
and can only be broken in the presence of substructure on scales
smaller than the image separation.  Hence, if the smooth model 
provides an adequate description of
the lens galaxy, one would expect  $R_{\rm cusp} \sim 0$ for these
lenses. This is clearly not the case and for this reason 
it is difficult to explain their
flux ratios using simple, smooth models.

If we make a comparison with simulations, one can see that the
pattern in Fig.~\ref{fig:cusprel}d clearly does not explain the observed
$R_{\rm cusp}$  of these four lenses. On the other hand,
substructure on few $\rm kpc$ scales and below provides enough
perturbations to $R_{\rm cusp}$ to explain the observed values.

The question arises, however, whether from the value of $R_{\rm cusp}$
we can conclude that 
we indeed see the effects of
CDM substructure.
\citet{keeton02} indeed argue that the cusp relation alone does not reveal
anomalous flux ratios in B1422$+$231. Still, detailed modelling for this
system shows that the flux ratios are anomalous. The
difficulties in modelling B1422$+$231 are not only a consequence of
a violation in cusp relation, but also that image D is a fainter
than predicted from smooth models. On the other
hand, the simulated disk galaxy shows violation of $R_{\rm
cusp}$ even though there are no clear mass clumps present in the
region where images are formed. Hence one has to be cautious when
making conclusions about the presence of CDM substructure based on the
value of $R_{\rm cusp}$ alone.

\subsection{The influence of noise on the cusp relation
\label{sc:cusprel_noise}}

The simulated cusp-relation can be reliably compared
with observations only if we know the noise properties of our
simulations. The effects of noise and
physical substructures  need to be disentangled through a 
detailed 
analysis.

\begin{figure}[ht!]
\begin{minipage}{8.5cm}
\begin{center}
\includegraphics[width=0.9\textwidth]{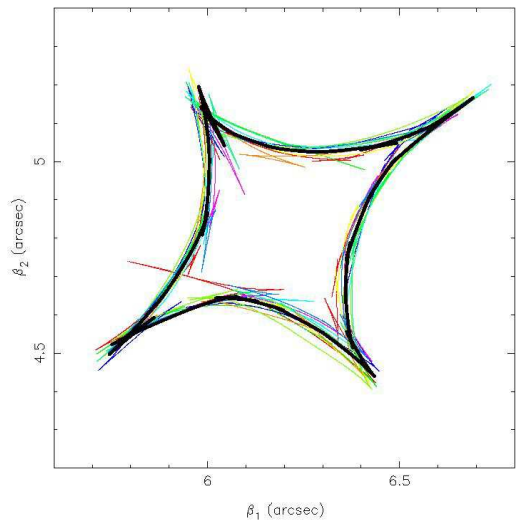}
\end{center}
\end{minipage}
\caption{The original caustic curve (thick line) of the halo from
Fig.~\ref{fig:cusprel}a and the corresponding caustic curves from the
ten bootstrapped maps plotted on top (thin lines).} 
\label{fig:caustic_noise}
\end{figure}

From the bootstrapping procedure (Sect.~\ref{sc:delaunay_noise}) we
also get an estimate of the error
in the cusp relation. We 
estimated the noise on $R_{\rm cusp}$ using a similar technique as for 
$\kappa$ in (\ref{eq:var}).
Moreover, we do not perform the analysis directly in the source plane
by subtracting the maps pixel-by-pixel.     
The problem is that bootstrapping somewhat changes the 
shape of the caustic curve (see also
Fig.~\ref{fig:caustic_noise}). Since we never observe the source plane
directly, in reality we can not distinguish between two shifted, but identical
caustics. We therefore have to match different bootstrap
maps in the image plane. We compare the image positions
generated by each source in the original frame with those generated by
bootstrapped lenses. Thus for each source position in the original map 
(see Fig.~\ref{fig:cusprel}a), we search for the source
position in the bootstrapped map such that the four image positions from both
maps differ as little as possible.

\begin{figure}[ht!]
\begin{minipage}{8.5cm}
\begin{center}
\includegraphics[width=0.9\textwidth]{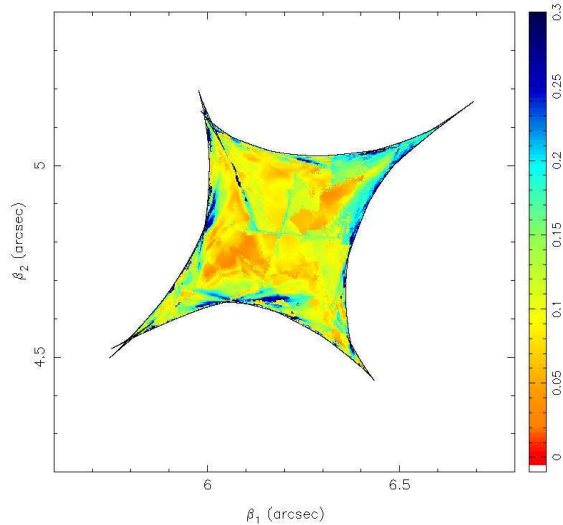}
\end{center}
\end{minipage}
\caption{The estimated absolute uncertainty $\sigma(R_{{\rm cusp},i})$ of
the cusp relation, calculated using the bootstrap analysis described in
Sect.~\ref{sc:delaunay_noise}. Note that the colour coding has changed
as compared with the other figures in this paper. The errors plotted
for each source position were calculated  using the systems from 
bootstrapped maps having similar image positions as the system
generated by the original lens; i.e. from Fig.\ref{fig:cusprel}a (see text).} 
\label{fig:cusprel_noise}
\end{figure}

In principle, one can redo the ray-tracing and calculate 
image fluxes for the position
in the original map using the bootstrapped-lens properties. 
However this is not necessary,
since our grid in the source plane is fine enough and we only 
need an approximate error estimate.  Figure~\ref{fig:cusprel_noise} 
shows the estimated absolute error  
$\sigma(R_{{\rm cusp},i})$ for the elliptical halo from
Fig.~\ref{fig:cusprel}a. As described above, each value of 
$\sigma(R_{{\rm cusp},i})$, plotted in grey-scale, does
not correspond to the error for a source position, but refers to the
error of the system with image positions similar to the original
ones.

The absolute error becomes slightly larger in the regions of the swallow
tails. This is,
however, not the effect of substructure vanishing in individual
bootstrapped realisations. It is rather the effect of slight position
changes of individual clumps. If one looks at the individual caustic
curves of bootstrapped halos (see Fig.~\ref{fig:caustic_noise}) the
swallow-tails are present in all realisations, although they change their
positions slightly. Since this hardly affects the image positions 
we cannot perfectly match the source positions $i'$ with the original 
source position $i$ in these regions; thus we are overestimating the
true error.

We conclude that the values of $R_{\rm cusp}$ in the close proximity 
of the cusp can be as high as $R_{\rm cusp} = 0.6$, with the error of 
$\pm 0.1$ as determined from  the bootstrapping.   Smoothing the 
substructure on scales as large as $\sim 1\:{\rm kpc}$ does not remove 
this effect, but reduces it. This is expected, since smoothing changes 
the profile of substructures.

Finally, we investigate
how well we can sample the smooth mass distribution given the number of
particles in the original simulation. We take an ellipsoidal
power-law density profile following $\rho(r) \propto r^{-2.9}$. The
power-law index was chosen, to closely reproduce the surface
mass density $\kappa$ of the simulation which follows 
$\kappa(\theta) \propto \theta^{-1.9}$
in the vicinity of the critical curve (see Sect.~\ref{sc:profile} for 
more details). The number of particles we use is the 
same as in the original N-body simulation. To each particle we assign the
average mass of the original sample, leaving the total mass of
the lens galaxy unchanged. We sample the density profile 
using a rejection
method (e.g. \citealp{numrec_c}). The resulting
particle distribution was again adaptively smoothed using Delaunay 
tesselation and we
perform the lensing analysis as in all previous cases.

\begin{figure}[ht!]
\begin{minipage}{8.5cm}
\begin{center}
\includegraphics[width = 0.9\textwidth]{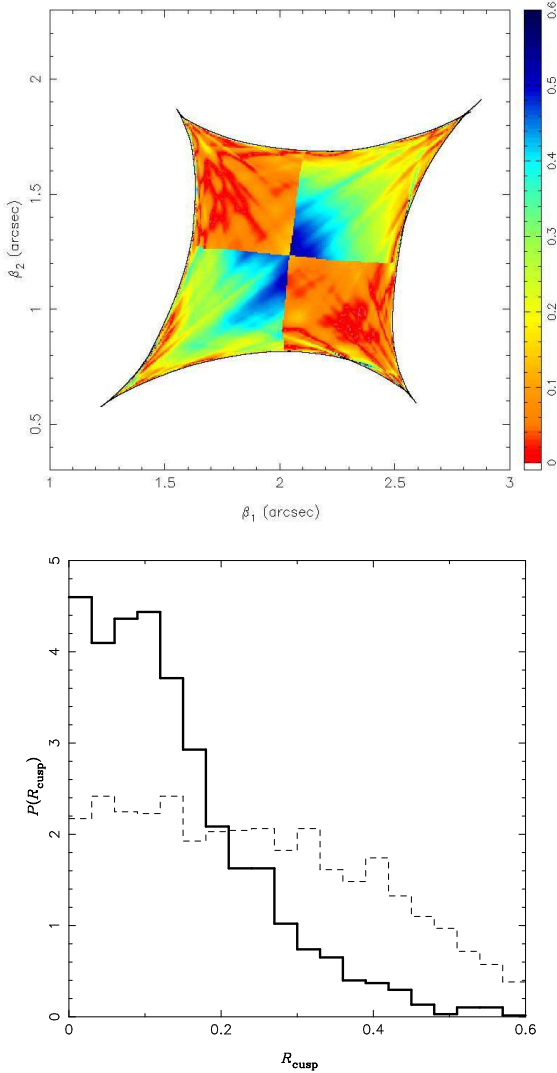}
\end{center}
\begin{center}
\includegraphics[width = 0.8\textwidth]{c4101_8b.ps}
\end{center}
\end{minipage}
\caption{{\bf(a)} The cusp relation for the smooth ellipsoidal model
sampled with the same amount of particles as present in 
the original N-body simulated elliptical galaxy. The redshifts of the
source and the lens were kept at  $z_{\rm l} = 0.81$ 
and  $z_{\rm s} = 3$. $R_{\rm cusp}$ is
plotted in gray-scale. On top we plot the corresponding 
caustic curve. The absence of large fluctuation in 
$R_{\rm cusp}$ as compared to
Fig.~\ref{fig:cusprel}a shows that the signal is not dominated by
the shot noise of the particles. {\bf (b)} The probability
distribution of $R_{\rm cusp}$ for systems with $\mu_{\rm tot} >
20$. The solid line gives the probability distribution for the sampled
smooth ellipsoidal model, while the dashed line corresponds to the
original N-body simulation (cf. Fig.~\ref{fig:cusprel}a).} 
\label{fig:cusprel_sample}
\end{figure}

The resulting cusp relation is given in 
Fig.~\ref{fig:cusprel_sample}a.  We have chosen the profile such  
that the particle densities
around the outer critical curves are similar in both
cases.  Only in that case can we reliably compare the noise properties of
$R_{\rm cusp}$. The fact that the caustic is larger than in the
case of the N-body simulated elliptical halo 
(compare to Fig.~\ref{fig:cusprel}a) is here of lesser
importance; it arises due the difference in the central
profile, far from the critical curve.

The absence of strong violations of $R_{\rm cusp}$ close to the caustic
in Fig.~\ref{fig:cusprel_sample} as compared to
Fig.~\ref{fig:cusprel}a confirms that deviation of $R_{\rm cusp}$ from
zero is not dominated by
the shot noise of the particles, but is due to genuine
substructure in the N-body simulation. For a more quantitative 
comparison we show the probability
distribution of $R_{\rm cusp}$ for systems with $\mu_{\rm tot} >
20$ in Fig.~\ref{fig:cusprel_sample}b as a solid line for the
sampled smooth halo and as a dashed line for the original. 
The much tighter distribution for the
sampled halo confirms the absence of strong violations of
$R_{\rm cusp}$ in the sampled smooth halo compared to the original 
simulation.

This analysis also shows the advantage in  
smoothing with Delaunay tesselation. If we only use a Gaussian kernel 
(of the same size as in Fig.~\ref{fig:cusprel}b), the deviations in the
cusp-relation for the sampled particle distribution are much larger.

\section{Lens models for synthetic image systems \label{sc:models}}

To explore the saddle point demagnification phenomena 
we fit the synthetic image systems using a singular isothermal
ellipse model with external shear (SIE+SH) \citep{ko94a}. We do not
include the flux ratios in the fit, since they are affected by the
substructure. Furthermore, we keep the lens position fixed. Using only
image positions we thus have 7 parameters and 8 constraints
(the parameters that we used are lens
strength, two components of the ellipticity of the lens, two external
shear components and the source position. The 4 image positions
provide 8 constraints). 

\subsection{Surface mass density profile \label{sc:profile}}
       
We find that the average unsigned magnifications predicted by the
SIE+SH model are higher than those from N-body simulations. This is
true for all systems generated with the same halo; a consequence of
the mass profile being steeper than isothermal around the typical
position where the images are formed.

We have calculated the $\kappa$ profile for the N-body simulated
elliptical by fitting the $\kappa$ with elliptical contours using the
{\tt IRAF.STSDAS} package {\tt ellipse}. For the inner $\sim 500 \:
{\rm pc}$ the profile is very close to isothermal with slope $-1.0 \pm
0.2$, the profile becomes steeper than isothermal with slope $-1.9 \pm
0.2$.  The break radius, where the profile becomes steeper than
isothermal, is smaller than the radius where images form.

We therefore over-predict the magnifications using an isothermal
profile. In order to deal with this problem, one would preferably use
a power-law profile with the index mentioned above for the lens
modelling, instead of SIE. However, these profiles require the
evaluations of hypergeometric functions and/or numerical integrations
\citep{grogin96,barkana98}. Modelling is therefore computationally too
expensive to apply to several 10\,000 sources. Moreover, this is not
critical, since we are not searching for the best fit macro-model, but
rather pretending that we observe the systems and try to fit them with
the model used for most observed lenses. Since we only observe the
flux ratios and not directly the magnifications, it is impossible to
compare magnification factors in practice. Thus one cannot spot the
difference in profiles using only this consideration when dealing with
real lens systems.

\subsection{Suppressed saddle points \label{sc:saddle}}
       
It was first noted by \citet{witt95} that the expected flux changes
due to stellar-mass perturbers differ between saddle points and minima
images. This was further investigated by \citet{schechter02} who
conclude that for fold images a stellar-mass component added to a
smooth mass distribution can cause a substantial probability for the
saddle point image to be demagnified w.r.t. the minimum.

Recently, \citet{kochanek03} and \citet{keeton03} investigated the
effect of singular isothermal sphere (SIS) mass clumps on the flux
perturbations for images of different types. Their conclusions are
similar; SIS perturbers also cause the brightest saddle point image to
behave statistically differently compared to those at minima.

Knowing whether the flux anomalies depend upon the image parity
for observed lenses would be a major step forward in identifying flux
anomalies either with substructure, or with propagation effects. 
Namely, if the observed flux anomalies depend upon
the image parity and its magnification we can set limits on the
influence of the ISM (see \citealt{kochanek03}).

\begin{figure}[ht!]
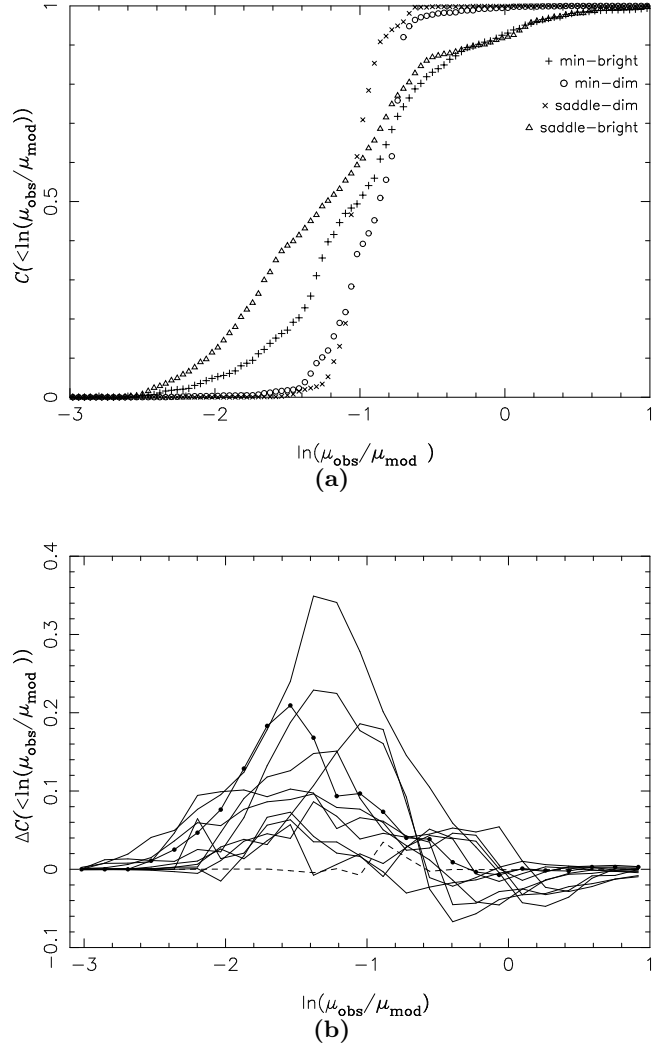

\begin{center}
\includegraphics[angle = -90,width=8.5cm]{c4101_9a.ps}\\
{\bf(a)}\\
\end{center}
\begin{center}
\includegraphics[angle = -90,width=8.5cm]{c4101_9b.ps}\\
{\bf(b)}\\
\end{center}
\caption{{\bf(a)} The cumulative flux residuals for each type of
image. Synthetic image systems were taken from the elliptical halo
(see also Fig.~\ref{fig:cusprel}a). Only image positions from the systems 
with  $\mu_{\rm tot} > 20$ generated from the elliptical halo 
were fitted using SIE+SH.
$\mu_{\rm
obs}$ are the magnification factors taken directly from N-body
simulations and  $\mu_{\rm mod}$ are the ones from best fit SIE+SH
models. 
{\bf(b)} The difference of 
the cumulative distribution of flux residuals  
$\Delta C\rund{<\ln(\mu_{\rm obs} / \mu_{\rm mod})}$ between the
  brightest saddle point and the brightest minimum images calculated using
  original halo (solid line with dots- dots also indicate the grid
  points used to evaluate $\Delta C$),
 bootstrapped images (solid lines) and halo when additionally  
smoothed on a scale of  $\sigma_{\rm G}\sim 5 \: {\rm kpc}$ (dashed line).} 
\label{fig:saddle_cum}
\end{figure}

To test whether saddle point images are also suppressed in our
simulation we proceed as follows.  Following \citet{kochanek03} we
define $C\rund{<\ln(\mu_{\rm obs} / \mu_{\rm mod})}$ to be the
cumulative probability distribution of flux residuals, where $\mu_{\rm
  obs}$ is the ``observed'' magnification of a simulated image and
$\mu_{\rm mod}$ is the magnification predicted by the best-fitting
smooth SIE+SH model (as described above).  In
Fig~\ref{fig:saddle_cum}a we plot $C\rund{<\ln(\mu_{\rm obs} /
  \mu_{\rm mod})}$ for the systems with ``observed'' total unsigned
magnifications $\mu_{\rm tot} > 20$ for the simulated elliptical halo
(see also Fig.~\ref{fig:cusprel}a).  We chose systems that have highly
magnified images, because they are affected by substructure at most.
We repeat the whole procedure with the same halo, but smoothed on a
scale of $\sigma_{\rm G}\sim 5 \: {\rm kpc}$; in this case essentially
no difference is seen among images of different parity, as expected.

For the original halo the cumulative distribution is much broader for
the brighter minimum and saddle point. This is in accordance with the
conclusions from \citet{mao98}; the higher the magnification, the more
is the image flux affected by substructure. Among the two most
magnified images, the saddle point is on average more demagnified
compared to the brighter minimum. We have examined two other
orthogonal projections of the mass density of the same halo and we
find that qualitatively the results do not change with projection.

The effect, however, is not as pronounced as in \citet{kochanek03}.
The reason is two-fold. First, our simulations have a resolution of
$\sim 10^8\:M_{\odot}$ and structure at this scale and below is
suppressed when using Delaunay tesselation. \citet{kochanek03},
however, used SIS clumps with masses of $10^6\:M_{\odot}$. Since these
are more numerous, they can enhance the effect. Further, fitting SIE
to the global mass profile is not fully justified. The mass profile is
known only for a handful of observed lenses. Whereas the lens galaxies
in MG1654$+$134, MG2016$+$112, 0047$-$281, and B1933$+$503 have a
nearly isothermal profile, the one of PG1115$+$080 seems to be steeper
(see \citealt{kochanek95,cohn01, treu02, treu02b, koopmans03}).
Besides the absence of substructure on scales $\lesssim
10^7\:M_{\odot}$, our synthetic systems and their modelled quantities
closely resemble the properties of realistic lenses and the way in
which these are modelled.

In the analysis of \citet{kochanek03} the synthetic images were
generated using an SIE macromodel with SIS substructure. This simplifies 
the model fitting and explains why they get a transition of cumulative 
distribution exactly at $\ln(\mu_{\rm obs} / \mu_{\rm mod})\sim 0$.

In addition, we have looked at the cumulative flux mismatch
distribution in the bootstrapped maps, to investigate the significance
of our results.  In the bootstrapped images we confirm the broader
distribution for bright minimum and saddle point images.  In
Fig~\ref{fig:saddle_cum}b we plot the difference of the cumulative
distribution of flux residuals $\Delta C\rund{<\ln(\mu_{\rm obs} /
  \mu_{\rm mod})}$ between the brightest saddle point and the
brightest minimum images for the original halo (solid line with dots),
bootstrapped images (solid lines) and halo when additionally smoothed
on a scale of $\sigma_{\rm G}\sim 5 \: {\rm kpc}$ (dashed line).
Positive values of $\Delta C$ thus denote the saddle point
demagnification. In all bootstrapped images $\Delta C$ is positive,
except for few points (corresponding to $C\rund{<\ln(\mu_{\rm obs} /
  \mu_{\rm mod})} \sim 1$). This confirms that the effect of the
saddle point demagnification is present and comparable to the original
halo, whereas in the smoothed halo this effect is not seen.  We
conclude that substructure on mass scales $\gtrsim 10^8\:M_{\odot}$
significantly contributes to the saddle point demagnification;
possibly going a long way in explaining the observed saddle point
demagnification.

\section{Conclusion \label{sc:conclusion}}
In this work we have studied strong gravitational lensing
properties of N-body simulated galaxies. In particular, we concentrated on
the influence of substructure on flux ratios on highly magnified
images. Such an analysis is crucial
in order to fully understand the lensing signal that we observe in realistic
lenses and to disentangle the influence on lensed-image fluxes due to 
propagation effects in the ISM and mass
substructure.

We have examined two strong lensing signatures of substructure, i.e. the
broken cusp relation observed in images that show a typical cusp
configuration and the saddle point suppression.
The saddle point suppression  has been previously studied using  
semi-analytic 
prescriptions of substructure \citep[e.g.][]{schechter02, kochanek03}. The
effect of substructure on the cusp relation, however, has up to now not been 
studied in detail.

In order to determine the magnitude of both effects we use
N-body simulated galaxies. The difference compared to the works of
\citet{schechter02} and \citet{kochanek03} is that we are using 
a representation of substructure that is as realistic as possible 
and do not make any
assumptions on the mass function and abundance of sub-halos. The
drawback, however, is the resolution of the simulations. We are
therefore not able to extrapolate the analysis to masses
$< 10^8\:M_{\odot}$. Still, the signatures of both effects are
clearly present, and in the future we plan to use higher-resolution
N-body plus gasdynamical  
simulations to explore their effects in greater detail.

The main question when dealing with N-body simulations is how much are
the magnification factors, that we use for synthetic image systems,
affected by noise which can mimic substructure of $ <
10^8\:M_{\odot}$. We show that the average relative noise in the
surface mass density $\sigma(\kappa) / \kappa$ lies below the $\sim 5
\%$ level for $\kappa \gtrsim 0.25$.  Second, in
Sect.~\ref{sc:cusprel_noise} we show that the results for $R_{\rm
  cusp}$ are not significantly affected by the noise, and are
dominated by physical substructure. The signal is dominated by several
resolved mass clumps, which in projection lie close to the Einstein
radius. Similarly, in the case of the phenomenon of suppressed saddle
points, the bootstrap analysis shows that the signal also here is not
dominated by noise.

The behaviour of $R_{\rm cusp}$ for sources close to a cusp is a very
promising tool to detect substructure. Its main advantage is that it
makes definite, model-independent predictions for the image
magnifications. These predictions can only be broken in the presence
of structure in the potential on scales smaller than the image
separation.  In Fig.~\ref{fig:cusprel}, where we used a simulated
elliptical galaxy to calculate $R_{\rm cusp}$, we clearly see these
effects.  When smoothing the substructure on larger scales we witness
the transition to the pattern that is common for generic smooth SIE
lens model.

However, the disc in the disk galaxy can also help to destroy the cusp
relation for sources in the vicinity of a cusp. We have calculated the
cusp relation pattern for the simulated disk galaxy, and even in the
absence of obvious substructure in the form of clumps we can see
strong violations of the cusp relation. This is expected, since the
edge-on disk gives $\kappa$-variations on the scales smaller than the
image separation and , similar to small-scale substructure.  One
cannot conclude from a broken cusp relation alone that we observe the
signatures of mass substructure in the form of clumps. However, the
observations show that most observed lenses are elliptical and
therefore one can concentrate on this morphology. Still, detailed
modelling is required in most cases (e.g. B1422$+$231) to clearly see
the effect.

The phenomenon of suppressed saddle points is a very strong prediction
that rules out a significant influence of the ISM on flux anomalies.
If flux anomalies depend on parity and magnification, they must
clearly be caused by lensing, if significant substructure is present.
Observations so far, show a clear parity dependence, which is more
obvious for highly magnified images.

Finally, our analysis shows that the two brighter images are more
affected by substructure than the two fainter ones. In addition, we
confirm that the brightest saddle point image in N-body simulated
systems has a higher probability to be demagnified, in accordance with
predictions from microlensing and from semi-analytic work by
\citep{kochanek03}. It is therefore necessary that all mass scales are
properly accounted for, in order to compare observations with
theoretical predictions in detail.

For future work, we plan to look for jet curvature as seen from N-body
simulations. At present, there is only one case of a curved jet
observed that is likely the cause of gravitational lensing
\citep{metcalf02b}. It will be interesting, however, to investigate
the probability of more of these occurrences. We plan to investigate
the signal one expects on average for multiple-imaged jets; this
signal is also less affected by noise in the simulation and low-mass
substructure.

In summary, gravitational lensing remains a very powerful tool for
testing the existence of CDM substructure. N-body simulated galaxies
do seem to produce the same effects as seen in observed lens systems.
In addition, systematics on flux anomalies (scatter broadening,
scintillation, microlensing) can be efficiently ruled out by
multi-frequency and higher-frequency observations of lenses.
Furthermore, the statistical analysis of large samples of lenses can
directly probe the properties of CDM substructure in galaxies to a
redshift of $z \sim 1$.  This provides a unique tool to measure the
evolution of these structures with cosmic time, as predicted in the
hierarchical structure formation scenario.

\begin{acknowledgements}
  We would like to thank to our referee for his constructive comments,
  Oleg Gnedin, Chris Kochanek and Oliver Czoske for many useful
  discussions that helped improve the paper, Vincent Eke for the
  initial conditions of the simulations, and Richard Porcas for
  careful reading of the manuscript.  This work was supported by the
  International Max Planck Research School for Radio and Infrared
  Astronomy at the University of Bonn, by the Bonn International
  Graduate School, by the Deutsche Forschungsgemeinschaft under the
  project SCHN 342/3--1, and by the NASA ATP program under grant NAG
  5-10827.
\end{acknowledgements}

\bibliography{/home/marusa/latex/inputs/bibliogr,/home/marusa/latex/inputs/bibliogr_cv}

\begin{thebibliography}{62}
\expandafter\ifx\csname natexlab\endcsname\relax\def\natexlab#1{#1}\fi

\bibitem[{{Abadi} {et~al.}(2003{\natexlab{a}}){Abadi}, {Navarro}, {Steinmetz},
  \& {Eke}}]{abadi02}
{Abadi}, M.~G., {Navarro}, J.~F., {Steinmetz}, M., \& {Eke}, V.~R.
  2003{\natexlab{a}}, \apj, 591, 499

\bibitem[{{Abadi} {et~al.}(2003{\natexlab{b}}){Abadi}, {Navarro}, {Steinmetz},
  \& {Eke}}]{abadi02b}
---. 2003{\natexlab{b}}, \apj, 597, 21

\bibitem[{{Barber} {et~al.}(1996){Barber}, {Dobkin}, \& {Huhdanpaa}}]{barber96}
{Barber}, B.~C., {Dobkin}, D.~P., \& {Huhdanpaa}, H. 1996, ACM Transactions on
  Mathematical Software, 22, 469

\bibitem[{{Barkana}(1998)}]{barkana98}
{Barkana}, R. 1998, \apj, 502, 531

\bibitem[{{Blandford} \& {Narayan}(1986)}]{blandford86}
{Blandford}, R. \& {Narayan}, R. 1986, \apj, 310, 568

\bibitem[{{Blandford}(1990)}]{blandford90}
{Blandford}, R.~D. 1990, \qjras, 31, 305

\bibitem[{{Blandford} \& {Kochanek}(1987)}]{blandford87}
{Blandford}, R.~D. \& {Kochanek}, C.~S. 1987, \apj, 321, 658

\bibitem[{{Brada{\v c}} {et~al.}(2002){Brada{\v c}}, {Schneider}, {Steinmetz},
  {Lombardi}, {King}, \& {Porcas}}]{bradac02}
{Brada{\v c}}, M., {Schneider}, P., {Steinmetz}, M., {et~al.} 2002, \aap, 388,
  373

\bibitem[{{Chen} {et~al.}(2003){Chen}, {Kravtsov}, \& {Keeton}}]{chen03}
{Chen}, J., {Kravtsov}, A.~V., \& {Keeton}, C.~R. 2003, \apj, 592, 24

\bibitem[{{Chiba}(2002)}]{chiba02}
{Chiba}, M. 2002, \apj, 565, 17

\bibitem[{{Cohn} {et~al.}(2001){Cohn}, {Kochanek}, {McLeod}, \&
  {Keeton}}]{cohn01}
{Cohn}, J.~D., {Kochanek}, C.~S., {McLeod}, B.~A., \& {Keeton}, C.~R. 2001,
  \apj, 554, 1216

\bibitem[{{Dalal} \& {Kochanek}(2002)}]{dalal02}
{Dalal}, N. \& {Kochanek}, C.~S. 2002, \apj, 572, 25

\bibitem[{{de Blok} \& {Bosma}(2002)}]{deBlok02}
{de Blok}, W.~J.~G. \& {Bosma}, A. 2002, \aap, 385, 816

\bibitem[{{Evans} \& {Hunter}(2002)}]{evans02b}
{Evans}, N.~W. \& {Hunter}, C. 2002, \apj, 575, 68

\bibitem[{{Evans} \& {Witt}(2003)}]{evans02}
{Evans}, N.~W. \& {Witt}, H.~J. 2003, \mnras, 345, 1351

\bibitem[{{Frigo} \& {Johnson}(1998)}]{frigo98}
{Frigo}, M. \& {Johnson}, S. 1998, in {1998 ICASSP conference proceedings},
  Vol.~3, 1381

\bibitem[{{Gaudi} \& {Petters}(2002)}]{gaudi02}
{Gaudi}, B.~S. \& {Petters}, A.~O. 2002, \apj, 574, 970

\bibitem[{{Grogin} \& {Narayan}(1996)}]{grogin96}
{Grogin}, N.~A. \& {Narayan}, R. 1996, \apj, 464, 92

\bibitem[{{Heyl} {et~al.}(1994){Heyl}, {Hernquist}, \& {Spergel}}]{heyl94}
{Heyl}, J.~S., {Hernquist}, L., \& {Spergel}, D.~N. 1994, \apj, 427, 165

\bibitem[{{Ibata} {et~al.}(1999){Ibata}, {Lewis}, {Irwin}, {Leh{\' a}r}, \&
  {Totten}}]{ibata99}
{Ibata}, R.~A., {Lewis}, G.~F., {Irwin}, M.~J., {Leh{\' a}r}, J., \& {Totten},
  E.~J. 1999, \aj, 118, 1922

\bibitem[{{Keeton}(2001)}]{keeton01}
{Keeton}, C. 2001, astro-ph/0112350

\bibitem[{{Keeton}(2003)}]{keeton03}
{Keeton}, C.~R. 2003, \apj, 584, 664

\bibitem[{{Keeton} {et~al.}(2003){Keeton}, {Gaudi}, \& {Petters}}]{keeton02}
{Keeton}, C.~R., {Gaudi}, B.~S., \& {Petters}, A.~O. 2003, \apj, 598, 138

\bibitem[{{Klypin} {et~al.}(1999){Klypin}, {Kravtsov}, {Valenzuela}, \&
  {Prada}}]{klypin99}
{Klypin}, A., {Kravtsov}, A.~V., {Valenzuela}, O., \& {Prada}, F. 1999, \apj,
  522, 82

\bibitem[{{Kochanek}(1995)}]{kochanek95}
{Kochanek}, C.~S. 1995, \apj, 445, 559

\bibitem[{{Kochanek} \& {Dalal}(2003)}]{kochanek03}
{Kochanek}, C.~S. \& {Dalal}, N. 2003, astro-ph/0302036

\bibitem[{{Koopmans} \& {de Bruyn}(2000)}]{koopmans00}
{Koopmans}, L. \& {de Bruyn}, A. 2000, \aap, 358, 793

\bibitem[{{Koopmans} {et~al.}(2003){Koopmans}, {Biggs}, {Blandford}, {Browne},
  {Jackson}, {Mao}, {Wilkinson}, {de Bruyn}, \& {Wambsganss}}]{koopmans03b}
{Koopmans}, L.~V.~E., {Biggs}, A., {Blandford}, R.~D., {et~al.} 2003, \apj,
  595, 712

\bibitem[{{Koopmans} \& {Treu}(2003)}]{koopmans03}
{Koopmans}, L.~V.~E. \& {Treu}, T. 2003, \apj, 583, 606

\bibitem[{{Kormann} {et~al.}(1994){Kormann}, {Schneider}, \&
  {Bartelmann}}]{ko94a}
{Kormann}, R., {Schneider}, P., \& {Bartelmann}, M. 1994, \aap, 284, 285

\bibitem[{{Kravtsov} {et~al.}(1998){Kravtsov}, {Klypin}, {Bullock}, \&
  {Primack}}]{kravtsov98}
{Kravtsov}, A.~V., {Klypin}, A.~A., {Bullock}, J.~S., \& {Primack}, J.~R. 1998,
  \apj, 502, 48

\bibitem[{{Lombardi} \& {Schneider}(2002)}]{lombardi02}
{Lombardi}, M. \& {Schneider}, P. 2002, \aap, 392, 1153

\bibitem[{{M{\" o}ller} {et~al.}(2003){M{\" o}ller}, {Hewett}, \&
  {Blain}}]{moeller02}
{M{\" o}ller}, O., {Hewett}, P., \& {Blain}, A.~W. 2003, \mnras, 345, 1

\bibitem[{{Mao}(1992)}]{ma92}
{Mao}, S. 1992, \apj, 389, 63

\bibitem[{{Mao} {et~al.}(2004){Mao}, {Jing}, {Ostriker}, \& {Weller}}]{mao04}
{Mao}, S., {Jing}, Y., {Ostriker}, J.~P., \& {Weller}, J. 2004, \apjl, 604, L5

\bibitem[{{Mao} \& {Schneider}(1998)}]{mao98}
{Mao}, S. \& {Schneider}, P. 1998, \mnras, 295, 587

\bibitem[{{Metcalf}(2002)}]{metcalf02b}
{Metcalf}, R. 2002, \apj, 580, 696

\bibitem[{{Metcalf} \& {Madau}(2001)}]{metcalf01}
{Metcalf}, R.~B. \& {Madau}, P. 2001, \apj, 563, 9

\bibitem[{{Metcalf} {et~al.}(2003){Metcalf}, {Moustakas}, {Bunker}, \&
  {Parry}}]{metcalf03}
{Metcalf}, R.~B., {Moustakas}, L.~A., {Bunker}, A.~J., \& {Parry}, I.~R. 2003,
  astro-ph/0309738

\bibitem[{{Metcalf} \& {Zhao}(2002)}]{metcalf02}
{Metcalf}, R.~B. \& {Zhao}, H. 2002, \apjl, 567, L5

\bibitem[{{Meza} {et~al.}(2003){Meza}, {Navarro}, {Steinmetz}, \&
  {Eke}}]{meza03}
{Meza}, A., {Navarro}, J.~F., {Steinmetz}, M., \& {Eke}, V.~R. 2003, \apj, 590,
  619

\bibitem[{{Moore}(1994)}]{moore94}
{Moore}, B. 1994, \nat, 370, 629

\bibitem[{{Moore} {et~al.}(1999){Moore}, {Ghigna}, {Governato}, {Lake},
  {Quinn}, {Stadel}, \& {Tozzi}}]{moore99}
{Moore}, B., {Ghigna}, S., {Governato}, F., {et~al.} 1999, \apjl, 524, L19

\bibitem[{{Navarro} \& {Steinmetz}(1997)}]{na97}
{Navarro}, J. \& {Steinmetz}, M. 1997, \apj, 478, 13

\bibitem[{{Press} {et~al.}(1992){Press}, {Teukolsky}, {Vetterling}, \&
  {Flannery}}]{numrec_c}
{Press}, W.~H., {Teukolsky}, S.~A., {Vetterling}, W.~T., \& {Flannery}, B.~P.
  1992, {Numerical recipes in C. The art of scientific computing} (Cambridge:
  University Press, 1992, 2nd ed.)

\bibitem[{{Quadri} {et~al.}(2003){Quadri}, {M{\" o}ller}, \&
  {Natarajan}}]{quadri02}
{Quadri}, R., {M{\" o}ller}, O., \& {Natarajan}, P. 2003, \apj, 597, 659

\bibitem[{{Schaap} \& {van de Weygaert}(2000)}]{schaap00}
{Schaap}, W.~E. \& {van de Weygaert}, R. 2000, \aap, 363, L29

\bibitem[{{Schechter} \& {Wambsganss}(2002)}]{schechter02}
{Schechter}, P.~L. \& {Wambsganss}, J. 2002, \apj, 580, 685

\bibitem[{{Schneider} {et~al.}(1992){Schneider}, {Ehlers}, \& {Falco}}]{sc92}
{Schneider}, P., {Ehlers}, J., \& {Falco}, E. 1992, {Gravitational Lenses}
  (Gravitational Lenses, Springer-Verlag Berlin Heidelberg New York.)

\bibitem[{{Schneider} \& {Weiss}(1992)}]{we92}
{Schneider}, P. \& {Weiss}, A. 1992, \aap, 260, 1

\bibitem[{{Springel} {et~al.}(2001){Springel}, {White}, {Tormen}, \&
  {Kauffmann}}]{springel01}
{Springel}, V., {White}, S.~D.~M., {Tormen}, G., \& {Kauffmann}, G. 2001,
  \mnras, 328, 726

\bibitem[{{Steinmetz}(1996)}]{st96}
{Steinmetz}, M. 1996, \mnras, 278, 1005

\bibitem[{{Steinmetz} \& {Navarro}(2000)}]{st00}
{Steinmetz}, M. \& {Navarro}, J. 2000, in ASP Conf. Ser. 197: Dynamics of
  Galaxies: from the Early Universe to the Present, eds. F. Combes, G. A.
  Mamon, and V. Charmandaris., 165

\bibitem[{{Steinmetz} \& {Navarro}(2003)}]{st01}
{Steinmetz}, M. \& {Navarro}, J.~F. 2003, New Astronomy, 8, 557

\bibitem[{{Swaters} {et~al.}(2000){Swaters}, {Madore}, \&
  {Trewhella}}]{swaters00}
{Swaters}, R.~A., {Madore}, B.~F., \& {Trewhella}, M. 2000, \apjl, 531, L107

\bibitem[{{Treu} \& {Koopmans}(2002{\natexlab{a}})}]{treu02}
{Treu}, T. \& {Koopmans}, L.~V.~E. 2002{\natexlab{a}}, \apj, 575, 87

\bibitem[{{Treu} \& {Koopmans}(2002{\natexlab{b}})}]{treu02b}
---. 2002{\natexlab{b}}, \mnras, 337, L6

\bibitem[{{Treu} \& {Koopmans}(2004)}]{treu04}
---. 2004, astro-ph/0401373

\bibitem[{{van den Bosch} \& {Swaters}(2001)}]{vdBosch01}
{van den Bosch}, F.~C. \& {Swaters}, R.~A. 2001, \mnras, 325, 1017

\bibitem[{{Wallington} \& {Narayan}(1993)}]{wa93}
{Wallington}, S. \& {Narayan}, R. 1993, \apj, 403, 517

\bibitem[{{Witt} {et~al.}(1995){Witt}, {Mao}, \& {Schechter}}]{witt95}
{Witt}, H.~J., {Mao}, S., \& {Schechter}, P.~L. 1995, \apj, 443, 18

\bibitem[{{Wo{\' z}niak} {et~al.}(2000){Wo{\' z}niak}, {Udalski}, {Szyma{\'
  n}ski}, {Kubiak}, {Pietrzy{\' n}ski}, {Soszy{\' n}ski}, \& {{\. Z}ebru{\'
  n}}}]{wozniak00}
{Wo{\' z}niak}, P.~R., {Udalski}, A., {Szyma{\' n}ski}, M., {et~al.} 2000,
  \apjl, 540, L65

\end{thebibliography}
\bibliographystyle{/home/marusa/latex/inputs/aa}
\end{document}